\documentclass[aps,twocolumn,floatfix,altaffilletter,superscriptaddress,preprintnumbers,tightenlines,showpacs,showkeys,notitlepage,nofootinbib]{revtex4-2}
\usepackage[T1]{fontenc}
\usepackage[colorlinks=true,citecolor=blue,linkcolor=blue]{hyperref}
\usepackage{braket,graphicx,physics,amsthm,amssymb,amsfonts,xcolor,booktabs,siunitx,mathrsfs}
\usepackage{mathtools}
\usepackage{slashed}
\usepackage[capitalise, english]{cleveref}

\usepackage[normalem]{ulem}
\usepackage{slashed}	
\usepackage{gensymb}
\usepackage{soul}
\allowdisplaybreaks  
\renewcommand{\vec}[1]{{\boldsymbol{#1}}}

\newcommand{\beq}{\begin{equation} }
\newcommand{\eeq}{\end{equation}} 
\newcommand{\bi}{\begin{itemize} }
\newcommand{\ei}{\end{itemize} }
\newcommand{\lp}{\left(}
\newcommand{\rp}{\right)}

\makeatletter
\providecommand*{\diff}%
	{\@ifnextchar^{\DIfF}{\DIfF^{}}}
\def\DIfF^#1{%
	\mathop{\mathrm{\mathstrut d}}%
		\nolimits^{#1}\gobblespace}
\def\gobblespace{%
	\futurelet\diffarg\opspace}
\def\opspace{%
	\let\DiffSpace\!%
	\ifx\diffarg(%
		\let\DiffSpace\relax
	\else
		\ifx\diffarg[%
			\let\DiffSpace\relax
		\else
  			\ifx\diffarg\{%
				\let\DiffSpace\relax
			\fi\fi\fi\DiffSpace}

\definecolor{Red}{rgb}{1.,0.,0.}

\usepackage{xspace}
\newcommand{\muthreee}{\textit{Mu3e}\xspace}

\definecolor{cborange}{HTML}{e69f00}
\definecolor{cbgreen}{HTML}{009e73}
\definecolor{cbyellow}{HTML}{f1dd42}
\definecolor{cblblue}{HTML}{56b4e9}
\definecolor{cbblue}{HTML}{0072b2}
\definecolor{defgrey}{HTML}{9f9f9f}
\definecolor{defgreen}{HTML}{8eba42}

\crefformat{equation}{(#2#1#3)}

\begin{document}

\preprint{}
\title{Displaced Searches for Axion-Like Particles and Heavy Neutral Leptons at Mu3e}

\author{Simon Knapen}
\affiliation{Berkeley Center for Theoretical Physics, University of California, Berkeley, CA 94720, U.S.A.}
\affiliation{Theory Group, Lawrence Berkeley National Laboratory, Berkeley, CA 94720, U.S.A.}

\author{Toby Opferkuch}
\affiliation{SISSA International School for Advanced Studies, Via Bonomea 265, 34136, Trieste, Italy}
\affiliation{INFN, Sezione di Trieste, Via Bonomea 265, 34136, Trieste, Italy}

\author{Diego Redigolo}
\affiliation{INFN, Sezione di Firenze Via G. Sansone 1, 50019 Sesto Fiorentino, Italy}

\author{Michele Tammaro}
\affiliation{INFN, Sezione di Firenze Via G. Sansone 1, 50019 Sesto Fiorentino, Italy}

\begin{abstract}
We present strategies for the \muthreee experiment to search for light, weakly coupled particles produced in rare muon decays, focusing on displaced $e^+e^-$ decays within the hollow target.
In most scenarios the backgrounds can be fully suppressed with a suitable set of cuts.
We furthermore quantify the interplay between displaced and prompt searches at \muthreee and existing constraints, showing how \muthreee has a unique opportunity to probe unexplored parameter space. 
\end{abstract}

%%%%%%%%%%%%%%%%%%%
% Main Section
%%%%%%%%%%%%%%%%%%%

\maketitle

\section{Introduction}
\label{sec:Introduction}

During the next decade we will see impressive advances in muon physics, which will complement the recent measurement of the muon anomalous magnetic dipole moment \cite{Muong-2:2023cdq}. 
Rare muon decay processes will be searched for in a series of upcoming experiments: COMET~\cite{COMET:2018auw}, DeeMe \cite{Teshima:2019orf} and Mu2e \cite{Mu2e:2014fns} will search for $\mu\to e$ conversions in the electromagnetic fields of nuclei, while MEGII~\cite{MEGII:2018kmf} and \muthreee ~\cite{mu3e:2020gyw} will aim for $\mu\to e\gamma$ and $\mu\to3e$ decays, respectively. 
The rates of these processes are predicted to be unobservable in the Standard Model (SM), and as such they are an excellent testing ground for extensions of the Standard Model where heavy new physics violates the lepton flavor symmetries.

In recent years, it has become clear that this program can also search for light, weakly coupled particles that could be produced in rare muon decays. There are several reasons as to why this approach is powerful: \emph{i)} Modern muon factories can produce very large amounts of muons, which enables searches for extremely rare decays. 
\emph{ii)} The muon is an extremely narrow state, which implies that even extremely weakly coupled particles could still be produced with an appreciable branching ratio. 
\emph{iii)} These decays are also often very distinct from the Standard Model and instrumental backgrounds, allowing for good to excellent signal to background discrimination. 
\emph{iv)} Finally, from a theoretical perspective it is plausible that weakly coupled particles, should they exist, do not have flavor universal couplings. This means that searches in exotic muon decays are complementary to probes that rely on the electron coupling, e.g.~beam dump or $e^+e^-$ collider experiments.

Concretely, it has been demonstrated that new light states with long lifetimes produced in lepton flavor-violating decay of the muon can be searched for in Mu2e \& COMET \cite{Shihua:thesis,Hill:2023dym,Xing:2022rob}, \muthreee  \cite{Perrevoort:2018ttp, Perrevoort:2018okj,Calibbi:2020jvd,Knapen:2023zgi} and MEGII \cite{Calibbi:2020jvd,Jho:2022snj}. Lepton flavor-conserving particles decaying promptly inside the detector can instead be probed by MUonE \cite{Galon:2022xcl} and \muthreee  \cite{Echenard:2014lma,Perrevoort:2018ttp,Hostert:2023gpk,Knapen:2023iwg}. 
In this work, we study the physics potential of a search for displaced decays of light dark sector states into $e^+e^-$ pairs  at \muthreee, which aims to collect $2.5\times10^{15}$ muon decays during phase-I, and $5\times10^{16}$ muon decays during phase-II. We suggest different analysis strategies and quantify their discovery potential. In particular, the interior of the hollow target provides a suitable fiducial volume for the displaced decays.

The most important feature in the signal topology is the presence (or absence) of missing momentum in the final state, e.g.~from neutrinos. 
To illustrate this, we consider three distinct decay topologies allowing us to sample a comprehensive set of signatures.
If there is no missing momentum, the invariant mass of the sum of all track momenta will reconstruct $m_\mu$.
This greatly reduces the SM background, similar to \muthreee's flagship $\mu \to 3e$ analysis \cite{mu3e:2020gyw}.
If missing momentum is present, the background is generically very large.
We will show however that in most cases it can still be very effectively suppressed. This relies on the dark sector particle being long-lived enough to decay inside the hollow target, as opposed to on the surface of the target.  

\begin{figure*}[t!]
\hspace{0 cm}
\begin{minipage}{0.25\textwidth}
\includegraphics[height=4cm]{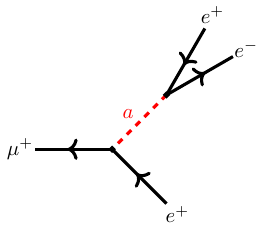}\\\vspace{-0.1cm} 
\end{minipage}\hspace{1cm}
\begin{minipage}{0.25\textwidth}
\includegraphics[height=4cm]{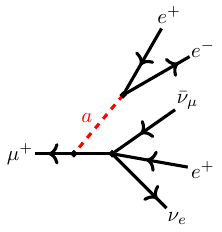}
\end{minipage}\hspace{1cm}
\begin{minipage}{0.25\textwidth}
\includegraphics[height=4cm]{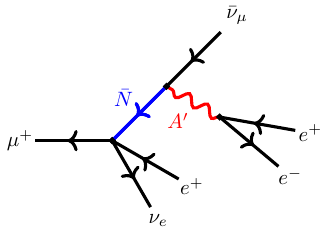}\vspace{-0.45cm}
\end{minipage}
\caption{\textbf{Left:} Two-body muon decay to an ALP produced through its lepton flavor violating coupling with the muon ($g_{\mu e}$ in \cref{eq:alpdefinition}) and decaying displaced to an $e^+ e^-$ pair. The invariant mass of the three tracks matches $m_\mu$. \textbf{Middle:} Four-body muon decay to a lepton flavor preserving ALP, produced mostly from its coupling with the muon $(g_{\mu\mu})$ and decaying displaced to an $e^+ e^-$ pair. The invariant mass of the three tracks does not match $m_\mu$, but the momentum vector of $a$ points back to the decay vertex of the muon. \textbf{Right:} Heavy neutral lepton decay through a dark photon. The invariant mass of the three tracks does not match $m_\mu$. If the $N$ is long-lived, the momentum of the $A'$ does not point back to the decay vertex of the muon. \label{fig:feyn}}
\end{figure*}

This paper is organized as follows: in \cref{sec:Models} we introduce our benchmark models and discuss their main features, in \cref{sec:analysis} we illustrate our search strategies and  estimate the expected backgrounds, in \cref{sec:Results}  we present our results and in \cref{sec:Conclusions} we conclude.

\section{Production channels and Benchmark Models}
\label{sec:Models}
We benchmark the expected sensitivity of \muthreee using two models:  
\begin{itemize}
\item A leptophilic Axion-Like Particle (ALP) decaying displaced into a $e^+e^-$ pair, first studied in Ref.~\cite{Heeck:2017xmg}, which can be produced in either two-body or four-body muon decays (see left and center diagram in \cref{fig:feyn}), \item A Heavy Neutral Lepton (HNL) decaying into a SM neutrino and a dark photon with the latter then decaying into an $e^+e^-$ pair (see far right diagram in \cref{fig:feyn}). In this model, first considered in Ref.~\cite{Hostert:2023gpk}, the long-lived particle can be either the HNL or the dark photon. 
\end{itemize}
 In both these models we compare the sensitivity of a search for displaced vertices with existing bounds from beam dump experiments, meson factories, supernova cooling and Big Bang Nucleosynthesis (BBN), and quantify how it complements the already planned searches at \muthreee for promptly decaying particles and invisible particles.

\subsection{Two-body and four-body muon decays: leptophilic ALPs}\label{subsec:LFVScalar}
We consider an ALP $a$ with  the following couplings to the lightest leptons 
\beq \label{eq:alpdefinition}
{\cal L}_a =  g_{\mu\mu} ia \bar\mu\gamma_5 \mu+g_{ee} ia \bar e\gamma_5 e + g_{\mu e} ia \bar \mu\gamma_5 e \,,
\eeq
where we only consider pseudoscalar couplings to the leptons, which are the ones of pseudo-Goldstone bosons or ALPs associated with approximate spontaneously broken abelian symmetries~\cite{Feng:1997tn,Calibbi:2020jvd}. For simplicity we focus on the  single helicity structure in \cref{eq:alpdefinition}, even though other structures for the flavor-violating fermion bilinear can occur in concrete ALP models~\cite{Feng:1997tn,Calibbi:2020jvd}.\footnote{The flavor-violating derivative coupling is in general \mbox{${\cal L}_a = 
 -\partial_\mu a \bar{\mu}\left[\gamma^\mu(g_{\mu e}^S+g_{\mu e}^A\gamma_5)\right]\!e$}, where $g_{\mu e}^A (m_\mu+m_e) \equiv g_{\mu e}$ is the coupling corresponding to the interaction in \cref{eq:alpdefinition} after integration by parts.} The different helicity structures will only affect the distribution of the polar angle at which $a$ is emitted, as the incoming muon beam at \muthreee is expected to be highly polarized.
This difference can be leveraged to improve the discrimination between signal and background in a prompt analysis \cite{Knapen:2023iwg}, but will only have a small impact on a search with little to no background. 

Independently of the production channel, we consider the $a\rightarrow e^+ e^-$ decay, assuming $m_a>2m_e$ (See left and middle diagrams in Fig.~\ref{fig:feyn}) which can be written as 
\beq
\Gamma\lp a\to ee \rp = \frac{g_{e e}^2 m_a}{8\pi}\left[1 - \frac{4 m_e^2}{m_a^2}\right]^{3/2} \simeq \frac{g_{e e}^2 m_a}{8\pi}\,,
\eeq
The decay length of $a$ in the lab frame can be computed as
\beq\label{eq:displacement}
\ell_a=(\gamma\beta)_ac\tau_a\,,
\eeq
where $(\gamma\beta)_a = |\vec{p}_a|/m_a $ is the particle boost factor, and $\tau_a = 1/\Gamma\lp a\to ee \rp$. $\ell_a$ can be approximated by
\beq
\ell_a \simeq  \SI{2}{\cm} \lp \frac{|\vec{p}_a|}{\SI{40}{\rm MeV}} \rp\lp \frac{10^{-5}}{g_{ee}} \rp^2  \lp \frac{\SI{10}{\MeV}}{m_a} \rp^2\, ,   
\eeq
where the ALP momentum, $|\vec{p}_a|$, depends on the ALP mass and the specific production channel. For instance, the momentum of a 30 MeV ALP produced in a two-body muon decays is $|\vec{p}_a| = 48.59\,{\rm MeV}$, while for the four-body muon decay the ALP tends to be softer, with an average momentum of $\langle |\vec{p}_a|\rangle\approx 28\,{\rm MeV}$ (see left panel of Fig.~\ref{fig:Pell-and-momenta-dists} in \cref{sec:mu3e_acceptance}). 

The coupling to muons will unavoidably generate a coupling to photons at one-loop, which leads to \cite{Shifman:1979eb,Bauer:2017ris}
\begin{align}
\Gamma\lp a\to \gamma\gamma \rp &\approx g_{\mu\mu}^2\frac{\alpha_{{\rm em}}^2}{9216\pi^3 }\frac{m_a^7}{m_\mu^6}.
\end{align}
We verified that this branching ratio is always negligible in the parameter space shown in the plots in \cref{sec:Results}, so we safely assume that the ALP decays to $e^+e^-$ pairs with branching ratio equal to 1, unless stated otherwise. 

We now discuss the two different production channels available for the leptophilic ALP of \cref{eq:alpdefinition}. 
First, the two-body decay production channel is controlled by the LFV coupling, $g_{\mu e}$. For $m_a < m_\mu-m_e$ the muon branching ratio to the ALP + electron final state is
\begin{align}
\label{eq:PW-mu-to-e-S}
{\rm BR}\lp\mu\to ea\rp &=  \frac{g_{\mu e}^2 m_\mu}{16\pi\Gamma_\mu} \left[1 - \frac{m_a^2}{m_\mu^2} \right]^2 \\
&\approx  7 \times 10^{-15} \left(\frac{g_{\mu e}}{10^{-15}}\right)^2 \left[1 - \frac{m_a^2}{m_\mu^2} \right]^2
\end{align}
where $\Gamma_\mu$ is the muon decay width.
With $2.5\times 10^{15}$ muons collected during phase I, this decay mode therefore remains detectable at \muthreee for a lepton-flavor violating coupling as tiny as $g_{\mu e}\approx 10^{-15}$. We will see this reflected in our results in Fig.~\ref{fig:LFVScalar:results} in \cref{sec:Results}.

Second, if we consider a model with only flavor-diagonal ALP couplings, the only open production channel is the four-body decay $\mu^+\to e^+\nu\bar{\nu} a$. 
Such a lepton-flavor conserving (LFC) ALP can either be attached to the $\mu^+$ or the $e^+$ leg, depending on the relative size of the $g_{\mu\mu}$ and $g_{ee}$ couplings (See middle diagram in Fig.~\ref{fig:feyn}). 
We focus here on a coupling regime where the production from the muon line dominates over the one from the electron line, which is guaranteed as long as e.g.~$g_{\mu\mu}\gg0.25g_{ee}$ for $m_a=\SI{30}{\MeV}$. 
The branching ratio for this process has been computed in Ref.~\cite{Knapen:2023iwg} and we refer to this work for the complete formulas. 
For $m_a\ll m_\mu $, this branching ratio can be approximated by
\begin{equation}
    {\rm BR}(\mu\to e\bar{\nu}\nu a)\approx \SI{3.8e-15}{}  \left(\frac{g_{\mu\mu}}{10^{-5}}\right)^2\ .
\end{equation} 
For $m_a\approx 30$ MeV, we estimate the maximal signal efficiency of \muthreee to be approximately 3.8\% (see \cref{sec:mu3e_acceptance}). 
In \cref{sec:analysis} we will moreover present a set of cuts which ought to reduce the background to negligible levels, without meaningfully impacting the signal efficiency.
This means that phase I should be able to probe couplings as low as $g_{\mu\mu}\approx 3\times 10^{-5}$, as reflected in Fig.~\ref{fig:LFC:results:geeVSgmumu} in \cref{sec:Results}.

\subsection{Three-body muon decay: heavy neutral leptons}
\label{subsec:HNL_DP}
We consider the model proposed in Ref.~\cite{Ballett:2019pyw}: the SM is enhanced by an extra $U(1)'$ gauge symmetry; the resulting vector gauge boson is the dark photon, $A_\mu'$. Additionally, we include two Heavy Neutral Leptons (HNL) states: $\chi$, neutral under SM and $U(1)'$, and $\psi$, charged under the new gauge group. Finally, an extra scalar $\Phi$ with the same charge as $\psi$ is present. 

The relevant part of the Lagrangian reads\footnote{In principle a mixed quartic $\lambda_{H\Phi} H^\dagger H \Phi^\dagger\Phi$ is allowed by the gauge symmetries of the model and will lead to a mixing between the SM Higgs and the dark singlet. We assume here that this quartic is small enough to neglect the Higgs portal couplings.}
\begin{align}\label{eq:HNLdef}
{\cal L} &\supset -\frac14 F_{\mu\nu}^\prime F^{\prime\,\mu\nu} -\frac{\epsilon}{2} F_{\mu\nu}' F^{\mu\nu}+ (D_\mu\Phi)^{\dagger}D^\mu\Phi \\
&+i\bar \chi \slashed\partial \chi + i\bar \psi {\slashed D} \psi-\left[\frac{\mu}{2} \bar{\chi}\chi^c+y_i \bar L_i \tilde H \chi^c + y_N \bar \chi \Phi \psi + {\rm h.c.}\right].\notag
\end{align}
Here, $\tilde{H}=i\sigma_2 H^\ast$, $F_{\mu\nu}$ ($F_{\mu\nu}^\prime$) is the photon ($A'$) field strength tensor, and $D_\mu = \partial_\mu - g^\prime A_\mu^\prime$ is the covariant derivative, with $g^\prime$ the $U(1)'$ gauge coupling. Furthermore, we allow for kinetic mixing of the photon and the $A'$, parametrized by the small $\epsilon$ term in \cref{eq:HNLdef}. 
The latter will generate the the $A'$ interactions terms with the SM fermions, with couplings proportional to $\epsilon Q_f$, where $Q_f$ is the fermion electric charge. 

A Majorana mass, $\mu$, for the HNL $\chi$ is allowed by the gauge symmetries of the model and breaks SM lepton number by two units. 
After both the new scalar singlet $\Phi$ and the SM Higgs take a VEV, $\langle \Phi\rangle=v_\Phi$ and $\langle \Phi\rangle=v$ respectively, the neutrino mass matrix from \cref{eq:HNLdef} is the same as the one of Inverse Seesaw Models~\cite{Mohapatra:1986aw,Mohapatra:1986bd} which has two HNL's $N$ and $N'$ plus three SM neutrinos.
Because the model has two HNLs, it allows for a portal between the HNL and the SM which does not break lepton number.
The corresponding coupling strength, $y_\alpha$ in this case, can therefore be large enough to give interesting signals at colliders, flavor factories and other accelerator-based experiments. 

In the inverse seesaw limit ($\mu\ll y_N v_\Phi< m_\mu$) the two HNL's are degenerate at a mass $\sqrt{2}m_N=y_N v_\Phi$. We label these mass eigenstates with $N_{1,2}$, which are maximal mixtures of the $\psi$ and $\chi$ fermions. 
Both have identical phenomenology, and we therefore the drop the subscript going forward and double the HNL production rate.
%\SK{I dropped the subscript in the equation below $N_i\to N$}
In the inverse seesaw limit, the mixing between the two sterile neutrinos and the SM neutrino of a given flavor $\alpha$ is moreover
\begin{equation}
    U_{\alpha N}=\frac{y_\alpha}{\sqrt{2}y_N}\frac{v}{v_\Phi}.
\end{equation}
For simplicity, we will assume here a flavor texture of $y_i$ is such that the HNL will mostly mix the muon neutrino; a more realistic flavor scenario including mixing with the electron neutrino will likely induce complementary phenomenological probes to the one discussed here. 

The HNL is produced from the three-body decay \mbox{$\mu^+\to e^+\nu_e N$} with differential rate \cite{Ema:2023buz} 
\begin{align}
    \frac{\diff \Gamma(\mu \to  e \,\nu_e N)}{\diff E_N} &= \frac{G_F^2 |U_{\mu N}|^2}{6\pi^3} |\vec{p}_N| \bigg[3 E_N \left(m_\mu^2 + m_N^2\right)\notag\\
    &-4 m_\mu E_N^2 - 2m_\mu m_N^2\bigg]\,,
\end{align}
where $|\vec{p}_N|$ is the absolute value of the outgoing HNL momentum, and $E_N$ is its total energy.

The decay width of the HNL and the dark photon can be written as~\cite{Ballett:2019pyw} 
 \begin{align} 
       & \Gamma(N\to\nu A') = \alpha_D|U_{\mu N}|^2 m_N \frac{(1 - r)^2}{4r} \lp r+\frac12  \rp \,,\\
       & \Gamma(A'\to e^+e^-)= \frac{\varepsilon^2 m_{A'}}{8\pi} \lp 1 - \frac{4 m_e^2}{m_{A'}^2} \rp\,,
    \end{align}
where we defined $\alpha_D = (g^\prime)^2/(4\pi)$ and $r\equiv m_{A'}^2/m_N^2$. 

We only consider the case where $N$ is long-lived, which is the case if $|U_{\mu N}|\ll 1$, while $\epsilon$ is chosen such that the $A'$ decays promptly. 
The phenomenology of the reverse case is qualitatively similar to the LFC ALP. 
The lab frame decay length of $N$ is
\beq
\ell_N\!\simeq \!0.8\,{\rm cm}\! \left( \frac{|\vec{p}_N|}{\SI{10}{MeV}} \right)\! \left(\frac{0.1}{\alpha_D} \right) \!\left(\frac{10^{-10}}{|U_{N\mu}|^2} \right) \!\left(\frac{\SI{10}{\MeV}}{m_{A^\prime}} \right)^2\!,
\eeq
in the limit $r\to0$.

\section{Analysis\label{sec:analysis}}

\subsection{The \muthreee  experiment}
The \muthreee experiment has a stopping target in the shape of a hollow, double cone with total length of \mbox{100 mm} and maximum diameter of 38 mm. 
The target is made of a thin Mylar sheet and is designed to minimize its impact on the tracks of the final state electrons and positrons. 
The tracking stations are build in close proximity to the target (see \cref{fig:mu3e}), leading to a very compact setup. 
This is needed to reconstruct soft tracks down to roughly \mbox{\SI{12}{\MeV}} momentum. 
The close proximity of the tracking layers to the target implies that the largest fiducial volume for displaced decays is the interior of the hollow target.

\subsection{Search strategies}

\begin{figure*}
\centering
    \includegraphics[width=0.48\linewidth,trim={1.4cm 0.3cm 3.3cm 1.7cm}, clip]
    {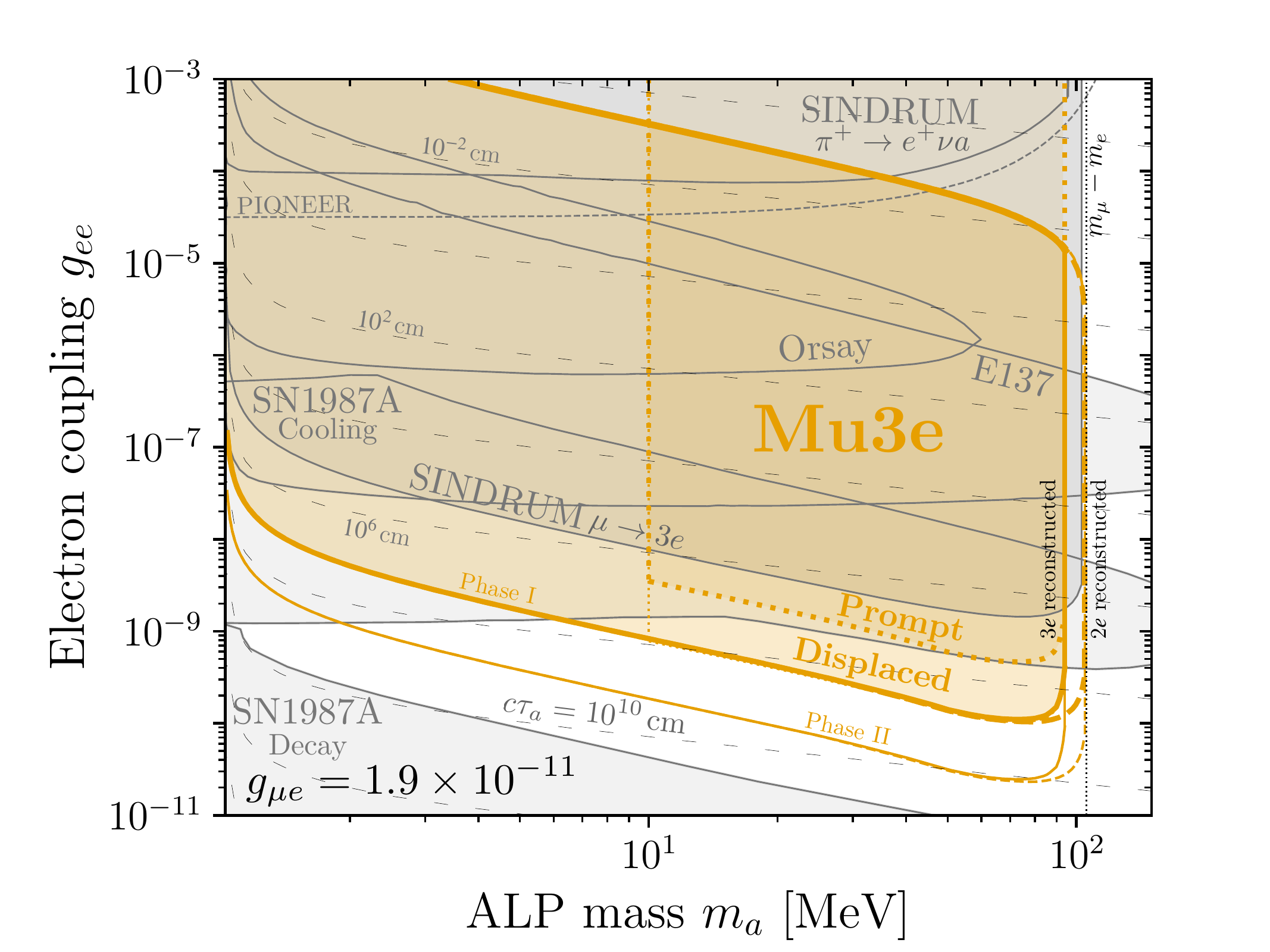}\hfill
    \includegraphics[width=0.48\linewidth,trim={1.4cm 0.3cm 3.3cm 1.7cm}, clip]{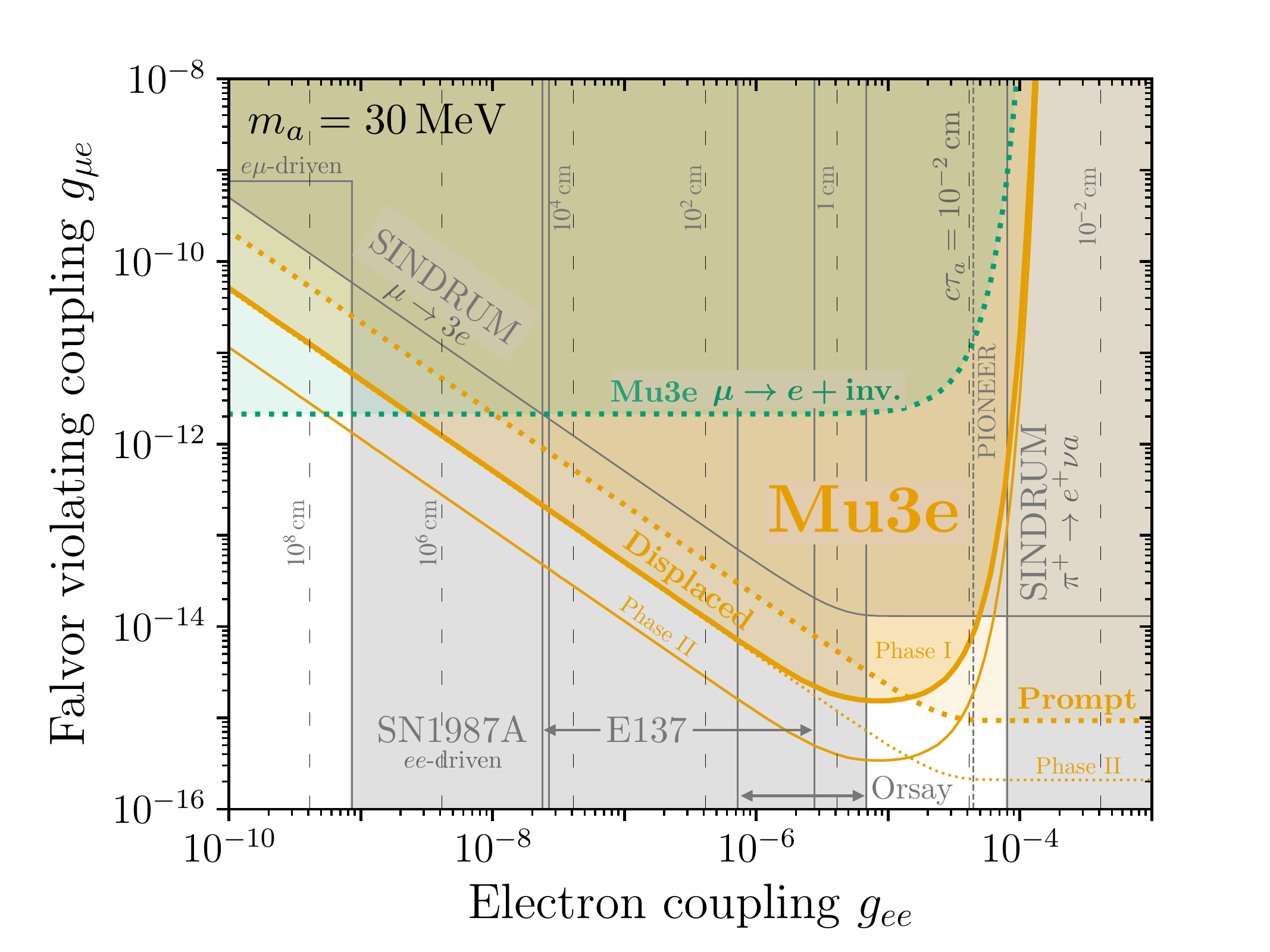}
    \caption{Projected 95\% confidence limits for the LFV ALP from searches for the $\mu^+\to e^+a, a\to e^+e^-$ at \muthreee Phase I (Phase II) with $\num{2.5e+15}$ ($\num{5e+16}$) muons decays. The prompt and displaced searches are indicated by the dotted and solid yellow curves. The dashed yellow curve indicates the reach of the displaced search in the part of parameter space where primary $e^+$ track is too soft to be reconstructed by \muthreee, see Appendix~\ref{sec:mu3e_acceptance} for details. For small the $g_{ee}$, \muthreee has sensitivity through the $\mu^+ \rightarrow e^+ + {\rm invisible}$ channel (phase I) (dashed green line) \cite{Perrevoort:2018okj}. The value $g_{\mu e}=\num{1.9e-11}$ was fixed in the left-hand panel, which saturates the existing $\mu^+ \rightarrow e^+ + {\rm invisible}$ bound from TWIST \cite{TWIST:2014ymv}. For clarity, this TWIST bound is omitted from the right-hand panel, as it is entirely covered by the current SINDRUM $\mu\to3e$ and future \muthreee $\mu^+ \rightarrow e^+ + {\rm invisible}$ bounds.
    The gray shaded regions indicate existing exclusions from electron beam dumps Orsay~\cite{Davier:1989wz} and E137~\cite{Bjorken:1988as} (as derived in \cite{Liu:2016qwd,Liu:2017htz}), SINDRUM~\cite{SINDRUM:1987nra,SINDRUM:1986klz} and SN1987a~\cite{ Ferreira:2022xlw}. We also show a projected bound from the proposed PIONEER experiment \cite{PIONEER:2022yag}.  The $\mu \to e\gamma$ bound is only relevant in a very small region when both $g_{\mu e}$ and $g_{ee}$ are large. This region is already covered by the SINDRUM $\pi\to e\nu a$ bound and we therefore omitted to $\mu \to e\gamma$ limit for clarity. The dashed gray contours indicate the proper decay length of $a$. 
    \label{fig:LFVScalar:results}
    }
\end{figure*}

Our baseline strategy is to require a reconstructed $e^-$ and $e^+$ track which form a vertex within the hollow target, at a distance sufficiently large to be distinct from the target's surface, reducing any overlap with prompt decays. 
We further require another $e^+$ track in the event which is consistent with being the decay product of a muon that was stopped on the surface of the target.\footnote{This implies that the signal efficiency drops to zero for $m_{a,A'}\gtrsim \SI{80}{\MeV}$, as detailed in \cref{sec:mu3e_acceptance}. For the $\SI{80}{\MeV}$ to $\SI{100}{\MeV}$ mass range we therefore also consider a signal region where we only require the tracks from the displaced vertex to be reconstructed. We verified that the loss in background suppression from not reconstructing the primary $e^+$ track is more than compensated for by the rapidly falling background distribution in $m_{ee}$ (see \cref{fig:radiativeBG}). The resulting projected sensitivity is indicated by the dashed lines in \cref{fig:LFVScalar:results}. }
In the remaining of the discussion we will refer to this track as the ``primary $e^+$ track'', i.e.~the direct daughter $e^+$ of the $\mu^+$.
Events where this primary $e^+$ track can be associated with the displaced vertex are vetoed, to suppress the background from muons decaying through internal conversion in the interior of the target (see \cref{sec:background}). 
The number of signal events can then be written schematically as 
\begin{equation}
    N_{{\rm sig}}= N_{\mu}\cdot {\rm BR}[\mu^+\to X+ e^+ (+{\rm inv.})]\cdot P (\ell_X)\cdot \epsilon \,,
\end{equation}
with $N_\mu$ the number of muons in the run and $X$ the long-lived particle, i.e.~$a$ or $N$ depending on the model. $P(\ell_X)$ is the probability that $X$ decayed within the hollow target and $\epsilon$ is the efficiency for all three tracks being reconstructed by \muthreee. We evaluate $P(\ell_X)$ and $\epsilon$ in \cref{sec:mu3e_acceptance}.

\begin{figure}[t]
\centering
\includegraphics[width=\linewidth]{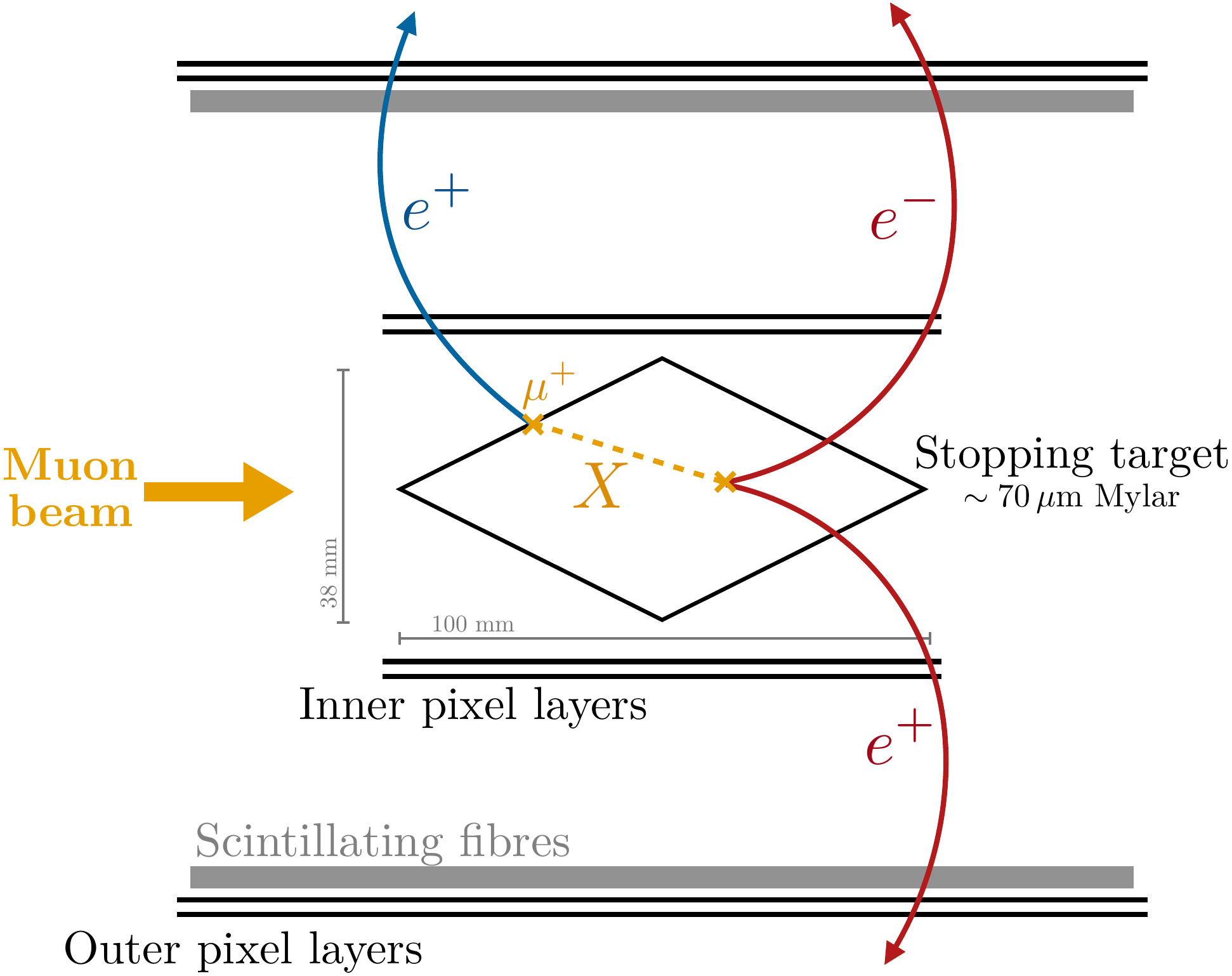}
\caption{Cross section of the inner part of the \muthreee detector with an example signal event, adapted from \cite{mu3e:2020gyw}. The $\mu^+$ is stopped on the surface of the target and decays to the primary $e^+$ track (blue track) and a long-lived new particle $X$ (yellow dashed line), which the decays within the hollow target to an $e^+e^-$ pair (red tracks). For clarity, only the central part of the \muthreee detector is shown.\label{fig:mu3e}}
\end{figure}

By requiring that the primary $e^+$ track is not associated with the displaced vertex, we eliminate all SM backgrounds for which all tracks originated from the same muon decay. 
Because of the high event rate, there are still considerable backgrounds from multiple coincident muon decays.
Before estimating these backgrounds in \cref{sec:background}, we introduce a few additional selection criteria which can help suppress them, at no cost to the signal efficiencies.
In particular, we can reconstruct the missing \mbox{four-momentum} vector by assuming that the muon decayed at rest
\begin{equation}
p^\mu_{\slashed{E}}\equiv P^\mu-\sum_i p^\mu_i
\end{equation}
with $p_i^\mu$ the four-momenta of the tracks and $P^\mu$ the four-momentum of the muon, \mbox{$P^\mu=(m_\mu,0,0,0)$}. For the LFV ALP (left-hand diagram in \cref{fig:feyn}) there are no invisible particles and we can therefore impose 
\begin{equation}\label{eq:scalarcut}
    p^\mu_{\slashed{E}} =0
\end{equation}
to within the momentum resolution of the experiment.  This restriction will be extremely powerful to reject coincidence backgrounds, as it is for the flagship $\mu\to 3e$ analysis \cite{mu3e:2020gyw}.  For the LFC ALP and the HNL models (middle and right-hand diagrams in \cref{fig:feyn}) the missing momentum is non-zero, but we can still impose the weaker condition that the missing momentum vector is time-like
\begin{equation}\label{eq:HNLcut}
    p^2_{\slashed{E}} >0.
\end{equation}
This condition is typically not satisfied if the the tracks in the event originate from two or more independent muon decays within the same timing window. It will therefore give us an additional handle for the LFC ALP and HNL models.

Finally, depending on the signal topology, we can impose a pointing cut. 
Concretely, we can demand that the reconstructed momentum vector of $X$ intersects the target surface in the same location as the primary $e^+$ track, within the spatial resolution of the detector (See \cref{fig:mu3e}). This condition can be imposed on the LVF scalar model and the HNL model if the $N$ decays promptly and the $A'$ decay is displaced. On the other hand, if the $N$ decays displaced, the momentum vector of the displaced $e^+ e^-$ pair will generally not point back to the origin of the primary $e^+$ track due to the presence of the neutrino in the final state. In this case we therefore cannot impose this cut.  

We will estimate the backgrounds for these various selections in the next section. 
For signal regions for which we expect negligible backgrounds, we approximate the expected $95\%$ exclusion bound by setting the signal rate to $N_{{\rm sig}}=3$ in the whole data sample. 
For signal regions for which we do expect background, we approximate the expected $95\%$ exclusion bound by setting the signal rate equal to $N_{{\rm sig}}=2.71\sqrt{N_\text{bkg}}$. 
For all further details on our simulation, detector modeling and cuts see \cref{sec:mu3e_acceptance}.

\subsection{Background estimates \label{sec:background}}
Displaced vertices can arise from the decays of muons that pass through the front cone of the target but are stopped in the gas within the hollow target. Assuming air at atmospheric pressure, we use the Bethe equation with inputs from the Particle Data Group \cite{ParticleDataGroup:2024cfk} to calculate the energy-dependent average energy loss and muon stopping power in air. For this purpose we assume the incoming muons have \SI{28}{\MeV} of momentum, the most probable momentum in the muon beam at PSI \cite{mu3e:2020gyw}. 
A little over half of all incoming muons are stopped on the front of the target. Using the spatial distribution of the stopped muons at the back of the target \cite{mu3e:2020gyw}, we compute the weighted average path length of a muon within the target. 
The ratio of the weighted average path length over the muon stopping range gives us the probability that a muon is stopped on the gas in the interior of the target. We then rescale this number to account for the fact that only 95\% of all muons are stopped by the target. 
Using this procedure, we estimate that 0.02 muons are stopped in the gas within the hollow target for every muon that is stopped on the stopping target's surface. The contribution of muons decaying in flight within the target is negligible in comparison.

\begin{figure}
    \centering
    \includegraphics[width=\linewidth,trim={1.4cm 0.3cm 3.4cm 0.7cm}, clip]{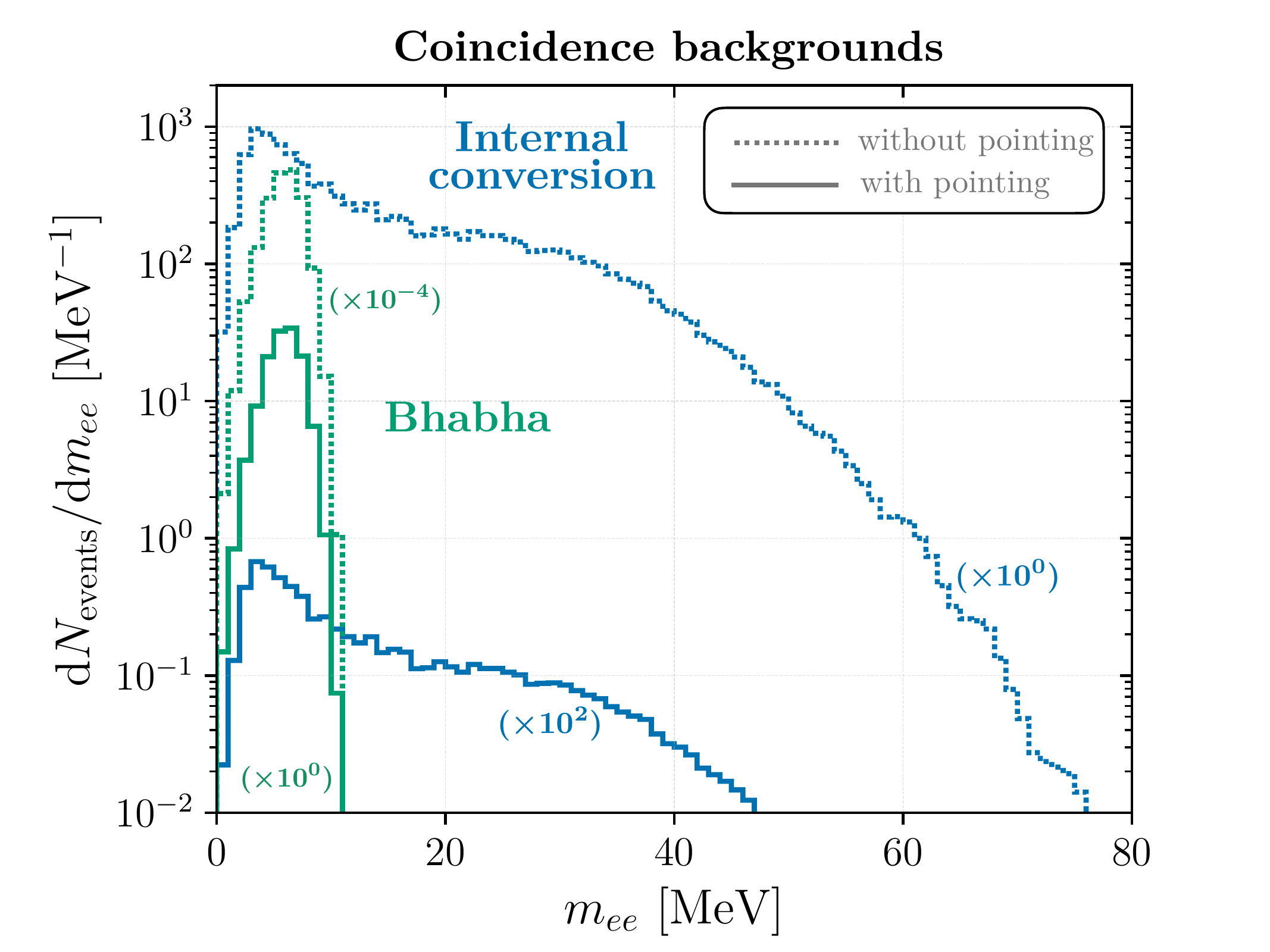}
    \caption{Estimated background from internal conversions and Bhabha scattering of muons stopped on the gas inside the target, with and without pointing the vertex back to the origin of the third track in the event. For clarity, we scaled the internal conversion background with pointing with a factor of $10^{2}$ and the Bhabha background without pointing with a factor of $10^{-4}$.}
    \label{fig:radiativeBG}
\end{figure}

The main physics background is therefore the internal conversion process \mbox{$\mu^+ \to e^+ e^- e^+ \nu\bar \nu$} of the muons that are stopped on air. We simulated  this process with \texttt{MadGraph5\_aMC@NLO 3.4.1} \cite{Alwall:2014hca}, using the reweighting procedure described in \cite{Knapen:2023iwg}. We demand that the electron and only one of the two positrons is reconstructed, which corresponds to a veto on vertices with three reconstructed tracks. 
For this background to mimic the signal topology in \cref{fig:mu3e}, we must further require the primary $e^+$ track is supplied by an unrelated Michel decay on the target surface, within the same timing window. For this purpose we assume a rate of $10^8$ muons per second and a 50 ps second timing resolution \cite{mu3e:2020gyw}. We further impose the missing momentum cuts in \cref{eq:scalarcut} for LFV ALP and \cref{eq:HNLcut} for the HNL model and the LFC ALP. For the LFV ALP, we find no residual background events. For the other two cases, this cut provides roughly one order of magnitude of background suppression.
The resulting spectrum is shown by the dashed blue line in Fig.~\ref{fig:radiativeBG} for a total sample of $\num{2.5e+15}$ muons stopped on the target.\footnote{In the HNL case the momentum distributions of the $e^+e^-$ pair also differ somewhat between this background and the HNL signal, in a manner which depends on the choices for $m_N$ and $m_{A'}$. It should therefore be possible to obtain a further $\mathcal{O}(1)$ improvement of the signal-to-background discrimination. We do not attempt to quantify this further improvement in this paper.}

For the LFC ALP, we can further require that the reconstructed momentum vector of the long-lived particle points back to the intersection of the primary $e^+$ track with the target surface. This provides an additional suppression factor of approximately $7\times 10^{-6}$ and renders this background completely negligible, as shown by the solid blue distribution in Fig.~\ref{fig:radiativeBG}. 

This suppression factor was estimated with a Monte Carlo simulation, assuming the displaced \mbox{$\mu^+ \to e^+ e^- e^+ \nu\bar \nu$} decays are distributed uniformly within the target and that the Michel decays on the target surface follow the simulated distribution in \cite{mu3e:2020gyw}. 
We furthermore assumed a spherically symmetric distribution for the reconstructed momentum vector of the long-lived particle, effectively neglecting the effect of the polarization in the muon decay amplitude. 
For the vertex resolution we assumed $\sigma_x=\sigma_y=0.5$ mm and $\sigma_z=0.3$ mm, which corresponds to the \muthreee's nominal alignment configuration \cite{mu3e:2020gyw}.

Bhabha scattering provides a second, more subtle background. 
Concretely, a muon stopped in the interior of the target can undergo a regular Michel decay, after which the positron can liberate an $e^-$ from a gas molecule. 
Because this electron is at rest initially, the maximum invariant mass of the $e^+e^-$ pair is $m_{e^+e^-}\leq \sqrt{m_e m_\mu}\approx 7\;\mathrm{MeV}$. 
The \muthreee collaboration estimates a total of roughly \mbox{$1.1\times 10^{11}$} Bhabha events in the target region of the detector \cite{mu3e:2020gyw}. 
We estimate the fraction of those events which occur inside the hollow target with the same procedure as for the internal conversions. 
We also require the presence of the primary $e^+$ track originating from an unrelated Michel decay on the surface of the target, and cut on the missing energy in the same manner as for the internal conversion background. 
This results in the dashed green histogram in Fig.~\ref{fig:radiativeBG}. 
We find that this background is very large for \mbox{$m_{e^+e^-}\lesssim 10$ MeV}, severely limiting the sensitivity if pointing is not available. 
For signals where the vertex points back to the origin of the primary $e^+$ track, this background is  small but non-zero (solid green histogram in \cref{fig:radiativeBG}). 
An important caveat is in order here: we assumed a gaussian smearing with mass resolution of \SI{1}{\MeV}, which we do not expect to be a reliable model for the background distribution several MeV beyond the \SI{7}{\MeV} edge. In other words, our simulation does not properly model the high $m_{e^+e^-}$ tail of the Bhabha background in the absence of a pointing cut. The \muthreee collaboration will however be able to measure this background and model it with a side-band analysis. 

In a sample with over $10^{15}$ events, one could also expect backgrounds due to the miss-reconstruction of pile-up events. For example, given the muon rate and timing resolution, one can expect $\sim 2 \times 10^{10}$ events with three simultaneous Michel decays. One of these muons can stop on the gas in the interior of the target and subsequently convert its $e^+$ to an $e^-$ through the Bhabha process. We estimate that there ought to be roughly $\sim 10^5$ such events. 
They are a priori not problematic, unless the track of the Bhabha $e^-$ accidentally intersects with the track from one of the Michel decays, leading to a spurious displaced vertex being reconstructed. 
A more sophisticated model of the \muthreee detector is needed to estimate this probability, and we do not attempt it here. We can however investigate the kinematic distributions of this background, while remaining agnostic about the suppression that can be achieved with vertex quality cuts.
We find that the invariant mass of both tracks peaks around $m_{e^+e^-}\approx \SI{30}{\MeV}$ and falls fairly fast for \mbox{$m_{e^+e^-}\gtrsim \SI{30}{\MeV}$}, indicating that it will likely always be very subdominant to the internal conversions once cuts on the vertex quality are imposed. 
If we further impose the pointing requirement, we expect this background to be completely negligible for all $m_{e^+e^-}$.

\section{Results}
\label{sec:Results}

The \muthreee  experiment aims to collect $2.5\times10^{15}$ muon decays during phase-I, and $5\times10^{16}$ during phase-II. The very large statistics, combined with excellent electron tracking resolution, will allow \muthreee to search for very rare muon decays into light, weakly coupled particles in several different decay topologies. 

\subsection{Lepton-flavor violating ALP}

\muthreee can search for lepton-flavor violating ALPs in several channels, whose comparative relevance depends on the lifetime of the ALP. 
First, if the ALP is either stable or long-lived relative to the size of the \muthreee detector, it will manifest itself as missing energy. 
\muthreee can search for this scenario in the $\mu\to e +a$ \cite{Perrevoort:2018ttp,Perrevoort:2018okj} or \mbox{$\mu\to 3e +a$} \cite{Knapen:2023zgi} channels, where one can leverage the phase space distributions to discriminate the signal from the Michel background. 
For larger $g_{ee}$ values, $a$ could decay promptly to an $e^+e^-$ pair. This scenario is a special case of the $\mu\to 3e$ flagship analysis, and is expected to be background free for $m_{a}\gtrsim \SI{10}{\MeV}$ \cite{mu3e:2020gyw}.

In \cref{sec:analysis} we have laid out a strategy to search for the intermediate regime, where the $a$ can be long-lived enough for its displaced decay vertex to be resolved separately by the \muthreee tracker. These three channels have overlapping sensitivity, highlighting the broad physics program of \muthreee .  This is shown in \cref{fig:LFVScalar:results}, where the solid thick (thin) yellow line indicates our estimate of the projected exclusion limit for the lepton flavor violating ALP from phase I (phase II) of \muthreee.
The dashed yellow line and the dashed green lines indicate $\muthreee$'s sensitivity in respectively the prompt and invisible channel. 

We see that a displaced search has some unique sensitivity, largely due to the much larger background for the invisible search. 
At the same time, the signal acceptance of \muthreee only drops as the inverse of the ALP lifetime, which can be seen clearly in the right-hand panel of Fig.~\ref{fig:LFVScalar:results}.
Quantitatively, we see that the displaced search is likely to outperform the invisible search for ALP proper decay lengths as large as $\sim 10^6$ cm. 
The signal acceptance of the prompt search is moreover suppressed if the ALP lifetime exceeds the spatial resolution of the detector.
As a result, the displaced search is expected to be the most powerful option for $10^{-9}\lesssim g_{ee}\lesssim 10^{-5}$.
Even in the regime where the displaced search is not the most sensitive, it would still be a logical and important next step in the event of an excess or discovery of signal events in either the prompt or invisible search.
Such a follow-up displaced search would then give \muthreee the opportunity to cross check the other analysis and measure the lifetime of the new particle in the event of a discovery.

To faithfully discuss the discovery potential of \muthreee, it is also important to map out how existing bounds constrain the same parameter space. We identify three classes of current experimental constraints: \emph{i)} past and current muon experiments, \emph{ii)} meson factories and beam dump experiments and \emph{iii)} astrophysical and cosmological bounds such as SN1987a and Big Bang Nucleosynthesis (BBN).

First, null results of searches for $\mu^+ \rightarrow e^+ + {\rm invisible}$ by Jodidio et al.~\cite{Jodidio:1986mz} and TWIST \cite{TWIST:2014ymv} provide a relevant limit on $g_{\mu e}$ as long as $a$ has an $\mathcal{O}(1)$ probability to decay outside the detectors. 
This is the case for most of values of the $g_{ee}$ we consider, and we have therefore set $g_{\mu e} \leq 1.9\times10^{-11}$ to saturate the TWIST bound in the left-hand panel of Fig.~\ref{fig:LFVScalar:results}. 
The $\mu \to 3e$ search at SINDRUM~\cite{SINDRUM:1987nra,SINDRUM:1986klz} constrains the parameter space at larger values for $g_{ee}$. 
At one loop level, the LVF ALP moreover induces the lepton-flavor violating decay $\mu\to e\gamma$, with branching ratio~\cite{Lavoura:2003xp, Heeck:2017xmg}
\begin{equation}
    {\rm BR}(\mu\to e\gamma)\sim 2.1\times10^9\times  (g_{ee}g_{\mu e})^2\,,
\end{equation} 
The combined limit from MEG and MEG II~\cite{MEG:2016leq,MEGII:2023ltw} is \mbox{${\rm BR}(\mu\to e\gamma)<3.1\times10^{-13}$}, which bounds the product \mbox{$g_{ee}g_{\mu e}\lesssim10^{-11}$}.

There are many other experiments which specifically probe the $g_{ee}$ coupling, and their complementarity with \muthreee is complicated and interesting.
These bounds are indicated by gray lines and regions.
Concretely, the SINDRUM experiment has placed a bound on the exotic pion decay $\pi^+\to e^+ \nu a, a \to e^+e^-$~\cite{SINDRUM:1986klz}, which is applicable for high $g_{ee}$ (short lifetime). 
The proposed PIONEER experiment is expected to improve on this bound \cite{PIONEER:2022yag}, but is likely to collect data after \muthreee.
The ALP can also be emitted through bremsstrahlung in electron beam dump experiments such as Orsay~\cite{Davier:1989wz} and E137~\cite{Bjorken:1988as}.
These experiments were sensitive to very small values of $g_{ee}$ and cover a significant part of the parameter space which would otherwise be available to \muthreee. 
We also explore the sensitivity from the decay of secondary muons produced in the NA62 target during beam dump mode, analogous to the muon bremsstrahlung process considered in \cite{Rella:2022len}. 
We find this to be very marginal with the existing NA62 dataset, though it could be competitive in possible future facilities such as HIKE and SHiP. We refer to Appendix~\ref{sec:OtherObservables} for details.

The anomalous cooling of a proto-neutron star, as SN1987A, excludes the region of couplings where the LFV scalar is efficiently produced in the star core, but is also able to escape~\cite{Calibbi:2020jvd, Carenza:2021pcm, Ferreira:2022xlw}. 
For very low $g_{ee}$ there is a second SN1987a bound from the decay of the ALP to $\gamma$-rays \cite{Ferreira:2022xlw}.
For completeness, we also mention that the ALP can provide a dangerous contribution to the number of effective neutrino species ($\Delta N_\text{eff}$), if it remains present after the neutrinos decouple in the early universe (see e.g.~\cite{Turner:1986tb,Brust:2013ova,Fradette:2017sdd,DEramo:2018vss,DEramo:2021usm}). 
For this $\Delta N_\text{eff}$ bound to apply, the ALP must have a lifetime $\gtrsim\SI{0.1}{\sec}$ and must have thermalized with the SM through processes which are suppressed by $g_{\mu e}$. 
On the right-hand panel of \cref{fig:LFVScalar:results}, it only applies in parts of parameter space which are already excluded by the SINDRUM $\mu\to3e$ search, and is therefore omitted.  
Finally, there likely exists a bound from Big Bang Nucleosynthesis from the late decay of a frozen-in subdominant relic density of ALPs. 
Quantifying such a bound is however outside the scope of this paper.

\subsection{Flavor conserving ALP}
By restricting the couplings of the light ALP to be flavor-conserving, $g_{ee}$ and $g_{\mu\mu}$, the kinematics of the signal channel drastically changes with respect to the LFV case. Production of $a$ can now happen uniquely via emission from the initial $\mu^+$, or from the final state $e^+$. The former (see middle diagram in Fig.~\ref{fig:feyn}) always dominates in the parameter space relevant for displaced decays of the ALP. The presence of neutrinos, which results in missing energy, also removes the possibility to reject backgrounds by reconstructing the muon mass from the total invariant mass $m_{eee}$. 
However, identification of a displaced vertex will still allow one to impose that the reconstructed $a$ momentum points back to the muon decay vertex, suppressing the background to a negligible level for \mbox{$m_a\gtrsim 10$ MeV} (see the green and blue solid lines in Fig.~\ref{fig:radiativeBG}).

\begin{figure}
\centering
    \includegraphics[width=\linewidth,trim={1.4cm 0.3cm 3.3cm 1.7cm}, clip]{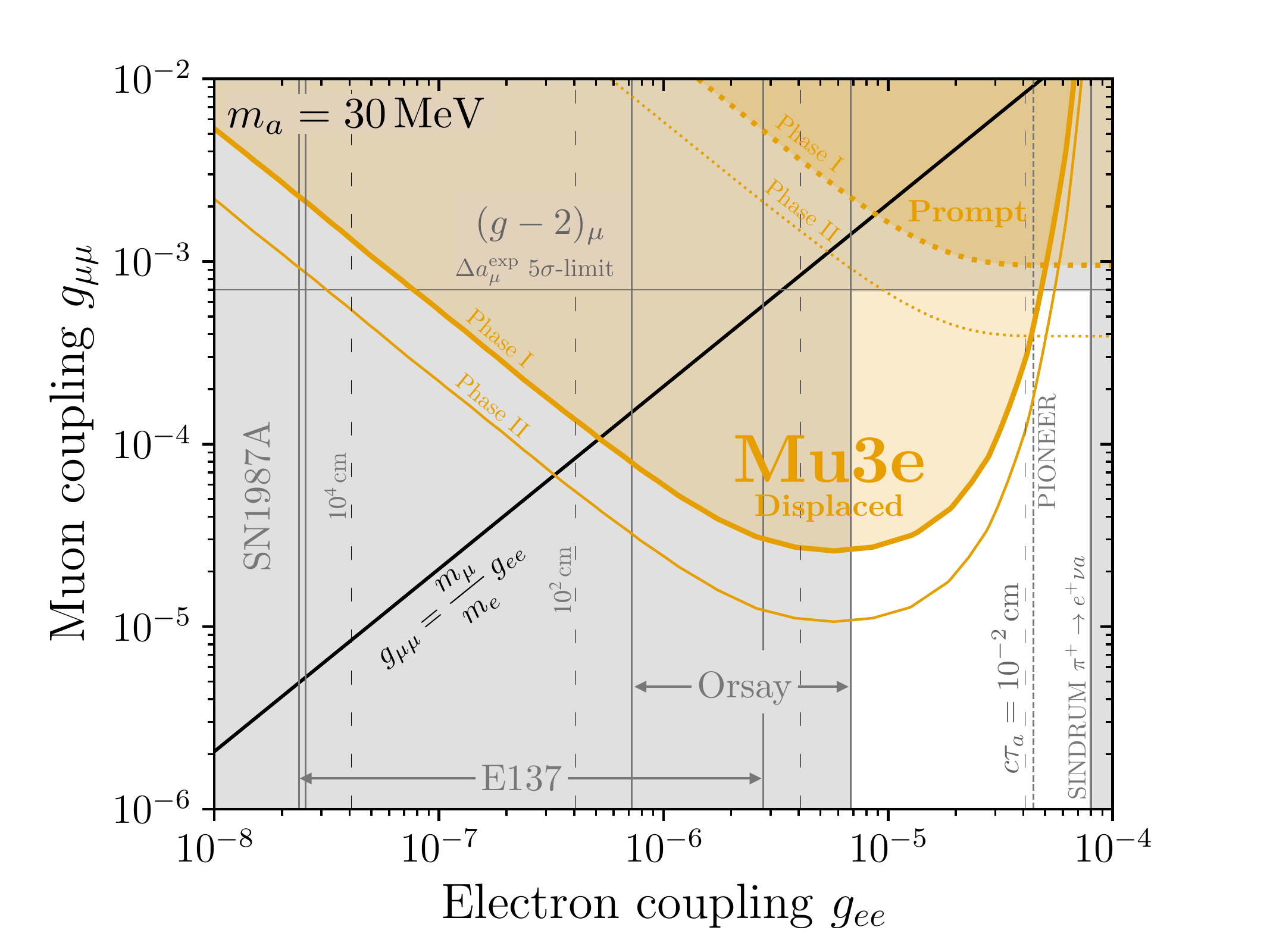}
    \caption{Projected limits on the leptophilic ALP scenario from searches of $\mu^+\to e^+\nu_e \bar\nu_\mu a, a\to e^+e^-$. We fix the mass of the ALP, $m_a = 30$ MeV, and show the reach as a function of the couplings to electrons, $g_{ee}$, and muons, $g_{\mu\mu}$. The solid black line indicates the benchmark of hierarchical couplings, $g_{\mu\mu} = g_{ee}m_\mu/m_e$. The bounds on $g_{ee}$ are the same as in Fig.~\ref{fig:LFVScalar:results}.  } 
    \label{fig:LFC:results:geeVSgmumu}
\end{figure}

To illustrate the sensitivity of a search for displaced vertices at \muthreee we fix $m_a = 30$ MeV in Fig.~\ref{fig:LFC:results:geeVSgmumu} and compare our results with present limits in the plane defined by $g_{ee}$ and $g_{\mu\mu}$. 
In a minimal UV completion of a lepto-philic ALP, one can expect a hierarchical structure of the couplings, proportional to the lepton yukawa couplings. A more general structure is however possible and therefore we allow both these couplings to be free parameters. We indicate the ``minimal'' expectation, namely $g_{\mu\mu} = g_{ee}m_\mu/m_e$, as a solid black line in Fig.~\ref{fig:LFC:results:geeVSgmumu}. Observables that depend specifically on the coupling to electrons have already been discussed in LFV case, and do not change here. Additionally, light ALPs that couple to muons can contribute to their magnetic anomalous moment. Given that there is currently still a tension between the lattice determination \cite{Borsanyi:2020mff} and the extraction from the R-ratio \cite{Aoyama:2020ynm}, it is not yet clear how to set an unambiguous limit. 
We choose to compare the experimental measurement \cite{Muong-2:2023cdq} with the most recent lattice determination \cite{Borsanyi:2020mff}, but as a very conservative bound we require a $5\sigma$ exclusion ($g_{\mu\mu}\lesssim7\times10^{-4}$), as opposed to the conventional 2$\sigma$ exclusion. This is large enough to subsume the difference between the two determinations. For more details, we refer to the discussion in \cite{Knapen:2023iwg}.

For large values of both $g_{ee}$ and $g_{\mu\mu}$, we expect high signal rates for prompt decays of $a$; the presence of the internal conversion background however severely limits the reach of \muthreee~\cite{Echenard:2014lma,Perrevoort:2018ttp,Knapen:2023iwg}; this is shown as a dashed yellow line. 
The solid yellow line show our estimate of \muthreee's reach from a search for displaced vertices. Thanks to the lower background levels, such a search would outperform a prompt search for $c\tau_a\gtrsim10^{-2}$ cm, and can reach $g_{\mu\mu}$ as low as $g_{\mu\mu}\sim10^{-5}$. 
Decay lengths up to $c\tau_a\sim10^3$ cm are presently covered by electron beam dump experiments, while even longer lifetimes are probed by stellar cooling.\footnote{For large $g_{\mu\mu}$ the supernova bound are modified due to trapping effects \cite{Bollig:2020xdr}.
In Fig.~\ref{fig:LFC:results:geeVSgmumu}, this effects is only present in the region which already excluded by the muon $g-2$ measurement, and we therefore neglect it. A bound from $W\to \mu \nu a$ and $W\to e \nu a$ decays also exists if the UV completion of \cref{eq:alpdefinition} contains addition sources of electroweak symmetry breaking \cite{Altmannshofer:2022ckw}.} 
Finally, we note that the Orsay bound disappears for $m_a\gtrsim \SI{50}{\MeV}$, and the E137 bound weakens as well. This means that $\muthreee$ could probe down to $g_{ee}\sim \num{e-6}$ or better for $\SI{50}{\MeV}\lesssim m_a \lesssim \SI{80}{\MeV}$.

\subsection{Heavy neutral leptons}
\begin{figure}
\centering
    \includegraphics[width=1.015\linewidth,trim={1.3cm 1.6cm 3.35cm 1.8cm}, clip]{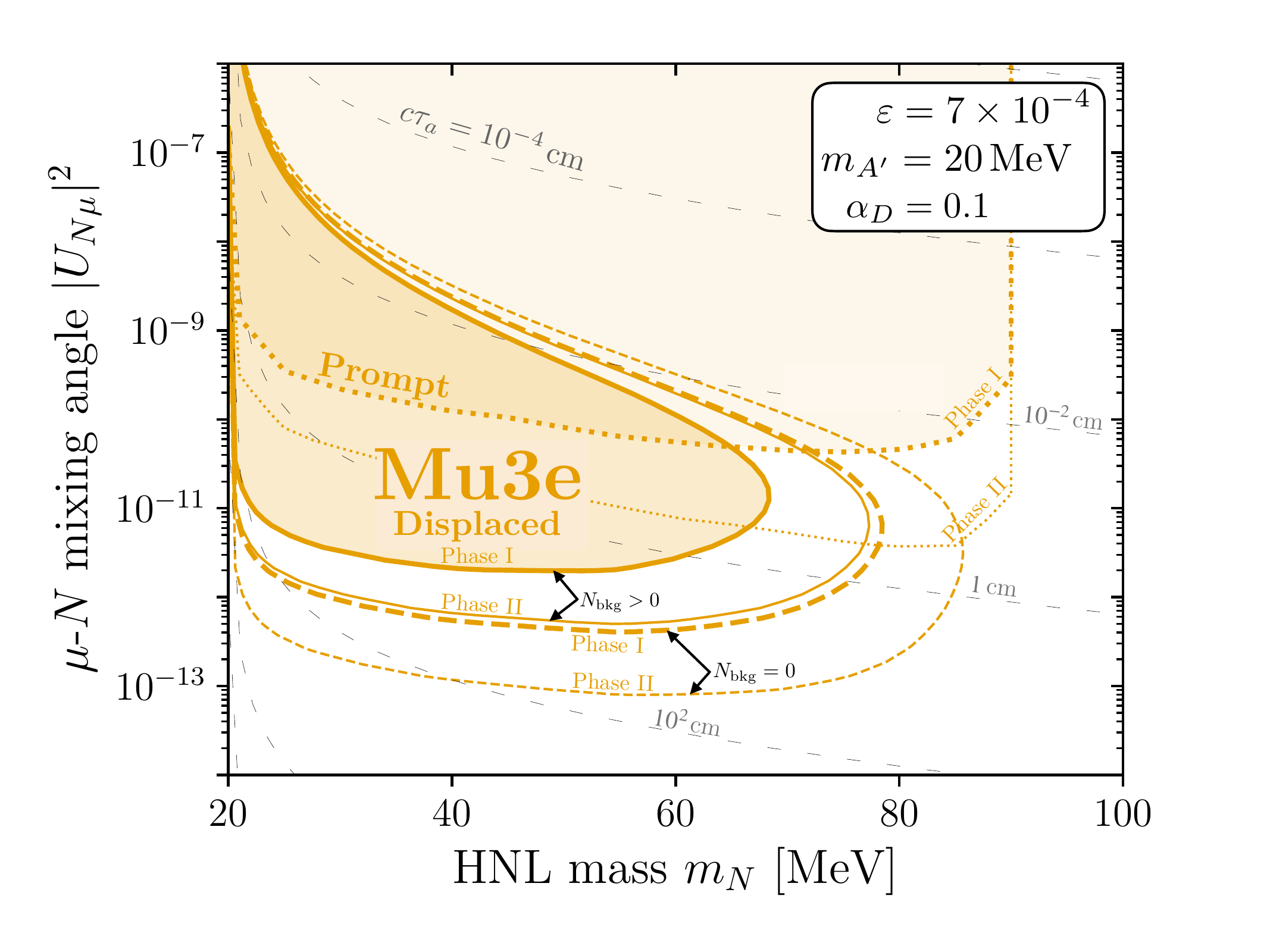}
    \caption{Projected limits on the HNL from searches for the $\mu^+\to e^+\nu_e N, N\to \bar\nu_\mu A^\prime, A^\prime\to e^+e^-$ at \muthreee. We scan the $m_N - |U_{N\mu}|^2$ plane, while we fix the remaining parameters, $\varepsilon$, $\alpha_D$ and $m_{A^\prime}$, as shown in the bottom right of the plot. This corresponds to the scenario where $N$ is displaced, while $A^\prime$ decays promptly. The yellow solid and dashed lines indicate the displaced search assuming no background events and our background estimate (see Sec.~\ref{sec:background} for details), respectively. The expected reach from prompt signals at \muthreee is shown as a dotted yellow line. The gray dashed contours indicate the proper decay length of $N$.}
    \label{fig:HNL:results:mN_VS_UN}
\end{figure}

As a final and experimentally most challenging example, we estimate sensitivity for the HNL model described in Sec.~\ref{subsec:HNL_DP}. The relevant channel at \muthreee is \mbox{$\mu^+\to e^+\nu_e N$}, followed by the \mbox{$N\to \bar\nu_\mu A^\prime$} and  \mbox{$A^\prime\to e^+e^-$} cascade, where \mbox{$2 m_e<m_{A^\prime}<m_N<m_\mu$}. We focus here on the case where $N$ is long-lived, while $A^\prime$ decays promptly. This scenario, unlike the previous ones, does not offer the possibility to suppress the backgrounds neither via matching the total invariant mass to the muon mass, nor by pointing of the reconstructed long-lived particle momenta to the muon decay vertex. 
Our estimate of the expected background is thus impacted by the internal conversion background (dashed blue line) in Fig.~\ref{fig:radiativeBG}. Additional considerations on the kinematical distributions of the tracks can help reduce this background; we leave this possibility for future studies.

As an example, we fix the $A^\prime$ mass to $m_{A^\prime} = 20$ MeV, the $U(1)^\prime$ gauge coupling to $\alpha_D = 0.1$, and the kinetic mixing with the SM photon as $\varepsilon = \num{7e-4}$. The latter ensures that the dark photon decays promptly on the typical distances of $\muthreee$, while still avoiding present constraints~\cite{Fabbrichesi:2020wbt}. The projected reach at \muthreee is then shown in Fig.~\ref{fig:HNL:results:mN_VS_UN}, where we distinguish two cases. Firstly, the dotted yellow line indicates the expected limit from a search for prompt decays, that is all three tracks are reconstructed to be originating from the muon decay vertex. This scenario is plagued by a very large background from the internal conversion process, $\mu^+\to e^+e^-e^+\nu\bar\nu$, with around $10^7$ events expected at $m_{ee} = 20$ MeV~\cite{Knapen:2023iwg}. The prompt search can thus probe at best $|U_{N\mu}|^2\sim6\times10^{-10}$, for masses where the detector efficiency is largest (see Fig.~\ref{fig:pTacceptance} in Appendix~\ref{sec:mu3e_acceptance}).

The solid yellow line shows our estimate of the limit \muthreee could set by searching for a displaced vertex. It includes our estimate for the background in Fig.~\ref{fig:radiativeBG}, which we don't expect to be negligible in this case. While the reach in $m_N$ is reduced, as larger masses receive a smaller boost from the muon decay, the reach in mixing angles is improved by almost two orders of magnitude. For completeness, the dashed yellow line shows what the reach would be if the expected background where to be negligible.

In minimal HNL models, the $m_N<m_\mu$ regime is already ruled out by a combination of the BBN bound \cite{Boyarsky:2020dzc} and searches for $\pi\to \mu N$ \cite{PIENU:2019usb} and $K\to \mu N$ \cite{Hayano:1982wu,Yamazaki:1984sj}\footnote{More recent searches such as \cite{NA62:2021bji} have excluded the low $m_N$ region from their signal selection.}, where $N$ behaves as missing momentum. 
By adding in a dark photon, Balett et.al.~\cite{Ballett:2019pyw} however drastically reduced the lifetime of the HNL, to the extent that neither set of bounds is universally applicable.  
For the parameter choices in Fig.~\ref{fig:HNL:results:mN_VS_UN}, this means that the HNL effectively decays promptly for $|U_{N\mu}|^2\gtrsim 10^{-8}$.
We can therefore instead consider the HNL's contribution to the $\pi^\pm \to \mu^\pm \nu_\mu e^+ e^-$ and $K^\pm \to \mu^\pm \nu_\mu e^+ e^-$ branching ratios.
The former does not appear to have been measured yet, while for the latter the existing measurements all excluded the part of the phase space for which $m_{e^+e^-}< 140$ MeV \cite{Poblaguev:2002ug,Diamant-Berger:1976oyg}, due to the backgrounds from processes such as $K^\pm \to \pi^\pm \pi^0$, followed by the Dalitz decay of the $\pi^0$. 
For this reason we consider the single-event sensitivity for NA62 in \cite{Ballett:2019pyw} to be too optimistic for the $m_{A'},m_N<m_\mu$ regime, and we did not include it in Fig.~\ref{fig:HNL:results:mN_VS_UN}.
To qualitatively gauge how such a measurement could impact the HNL parameter space, we can instead engage in a thought experiment, by assuming a hypothetical future measurement of the $m_{e^+e^-}<m_\mu$ part of the $K^\pm \to \mu^\pm \nu_\mu e^+ e^-$ phase space. 
The theoretical prediction for $K^\pm \to \mu^\pm \nu_\mu e^+ e^-$ branching ratio is $3.15\times 10^{-6}$ for $m_{ee}>20$ MeV \cite{Bijnens:1992en}.
If this could be measured with $\sim 10\%$ precision, this would correspond to excluding $|U_{N\mu}|^2\gtrsim 3\times 10^{-7}$  in Fig.~\ref{fig:HNL:results:mN_VS_UN} and be complementary to the sensitivity of \muthreee in the displaced channel.

\section{Conclusions}
\label{sec:Conclusions}
We have studied \muthreee's sensitivity to various dark sector models by searching for a displaced vertex within the interior of the stopping target. 
To most effectively reject the coincidence backgrounds, it is necessary to require a reconstructed track from the muon's decay vertex, in addition to the two tracks of the displaced vertex (see Fig.~\ref{fig:mu3e}).  
To cover all possible models and decay topologies (see Fig.~\ref{fig:feyn}), we recommend three complementary signal regions:
\begin{itemize}
    \item Require the three tracks to reconstruct $m_\mu$, and the momentum vector of displaced $e^+e^-$ pair to point to the intersection of the third track with the stopping target. This signal region is sensitive to lepton-flavor violating ALPs.
    \item Do not require the three tracks to reconstruct $m_\mu$, but require the momentum vector of displaced $e^+e^-$ pair to point to the intersection of the third track with the stopping target. This signal region is sensitive to lepton flavor conserving ALPs or any other light state radiated off the muon Michel decay and subsequently decaying displaced into $e^+e^-$ pairs. 
    \item Require three reconstructed tracks, without either of the above criteria. This signal region is sensitive to heavy neutral leptons with a dark photon decay portal.
\end{itemize}
In all three channels, a displaced search at \muthreee will provide complementary sensitivity to the already planned prompt and invisible searches, further rounding out the broad physics potential of the experiment.

\section*{Acknowledgements}
We are grateful to Kevin Langhoff, Ann-Kathrin Perrevoort and Jure Zupan for useful discussions.
The work of SK was supported by the Office of High Energy Physics of the U.S. Department of Energy under contract DE-AC02-05CH11231. We thank the Galileo Galilei Institute for hospitality during the development of this study. This work was performed in part at Aspen Center for Physics, which is supported by
National Science Foundation grant PHY-2210452. The work of DR is supported in part by the European Union - Next Generation EU through the PRIN2022 Grant n. 202289JEW4.
    
\bibliographystyle{JHEP}
\bibliography{bibliography.bib}

\appendix
\renewcommand{\onecolumngrid}{%
  \do@columngrid{one}{\@ne}
  \def\set@footnotewidth{\onecolumngrid}%
  \def\reset@font@note{\let\@makefnmark\@makefnmark@onecol}%
}
\onecolumngrid

%===========================================================================
\section{Detector modeling}
\label{sec:mu3e_acceptance}
%===========================================================================
Here we provide additional details on the \muthreee  signal acceptance and detector modeling. 
There are two main factors at play; \emph{i)} the geometric acceptance given by the shape of the stopping target and \emph{ii)} the kinematics of the muon decay products in conjunction with the detector's threshold on the $e^+$ and $e^-$ transverse momenta. 
To recover sensitivity for high $m_a$, it is imperative to also consider the case where only the tracks from the displaced vertex are being reconstructed by \muthreee.
The number of signal events is given by
\begin{equation}
    N_{{\rm sig}}= N_{\mu}\frac{1}{\Gamma_\mu}\int \diff^3\vec p_X \diff^3 \vec p_e \frac{d\Gamma[\mu^+\to X+ e^+ (+{\rm inv.)}]}{\diff^3 \vec p_X \diff^3\vec p_e} \cdot P(\vec p_X,c\tau_X)\cdot \epsilon_i (\{\vec p^i_e\})\,,\label{eq:mastereqappendix}
\end{equation}
with $\vec p_X$ and $\vec p^i_e$ the momenta of the $X$ particle and the final state tracks. $P(\vec p_X,c\tau_X)$ is the probability that $X$ decayed within the hollow target. While $\epsilon_i  (\{\vec p^i_e\})$ is the efficiency for all three tracks being reconstructed ($\epsilon_{3e} (\{\vec p^i_e\})$) or the efficiency of only the displaced tracks being reconstructed ($\epsilon_{2e} (\{\vec p^i_e\})$).
To compute $N_{{\rm sig}}$, we simulate the differential decay width with \texttt{MadGraph5\_aMC@NLO} \cite{Alwall:2014hca}, and then reweight the events by separately calculating $P(\vec p_X,c\tau_X)$ and $\epsilon_i (\{\vec p^i_e\})$.

\subsection{Geometric Acceptance}
The \muthreee  stopping target consists of two joined conical shapes, forming a double-cone structure aligned along the beam-line. The radius of the cones at their widest point is $r_c = \SI{19}{\mm}$, each one with aperture of $\alpha = 20.8\degree$ and height $h = \SI{50}{\mm}$. In the experiment's coordinate system~\cite{mu3e:2020gyw}, the base of the double cone is centered at $(0,0,0)$, with the $z$-axis chosen to be aligned with the two cone axes, while the $x$ and $y$ axes are chosen accordingly to form an orthonormal basis. The two tips of the target are at $(0,0,-50)$ and $(0,0,50)$ respectively. A point on the double cone is thus represented as $\mathbf{x}(z,\beta)=(r(z) \cos\beta, r(z)\sin\beta, z)$, where $r(z) = (h-|z|)\tan\alpha$, for $z\in[-50,50]$ mm and $\beta\in[0,2\pi]$.

On an decay-by-decay basis, we use $\mathbf{x}_0(z_0,\beta_0)$ to denote the location on the target where the muon stops. The distribution of stopped muons as function of $z_0$ was simulated in \cite{mu3e:2020gyw}. 
This allows us to define a probability distribution $p(z_0)$, which is independent on the azimuthal angle $\beta_0$ due to the cylindrical symmetry of the experimental setup. This distribution is small near the tip of both cones and peaks near $z_0\approx \pm 20$ mm.

\begin{figure}
\centering
    \includegraphics[width=0.495\linewidth,trim={1.4cm 0.6cm 2.9cm 1.3cm}, clip]{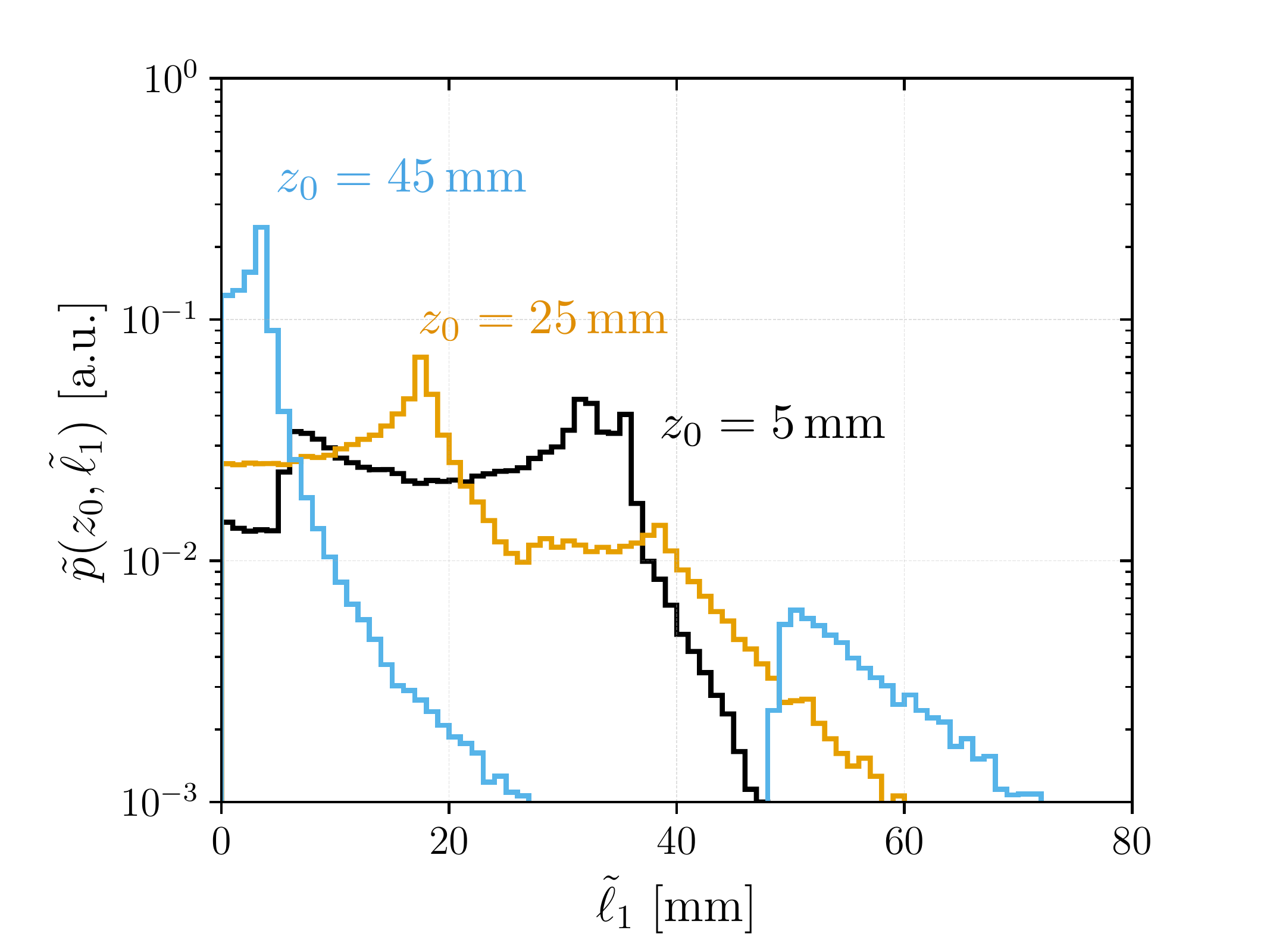} 
    \includegraphics[width=0.495\linewidth,trim={1.4cm 0.6cm 2.9cm 1.3cm}, clip]{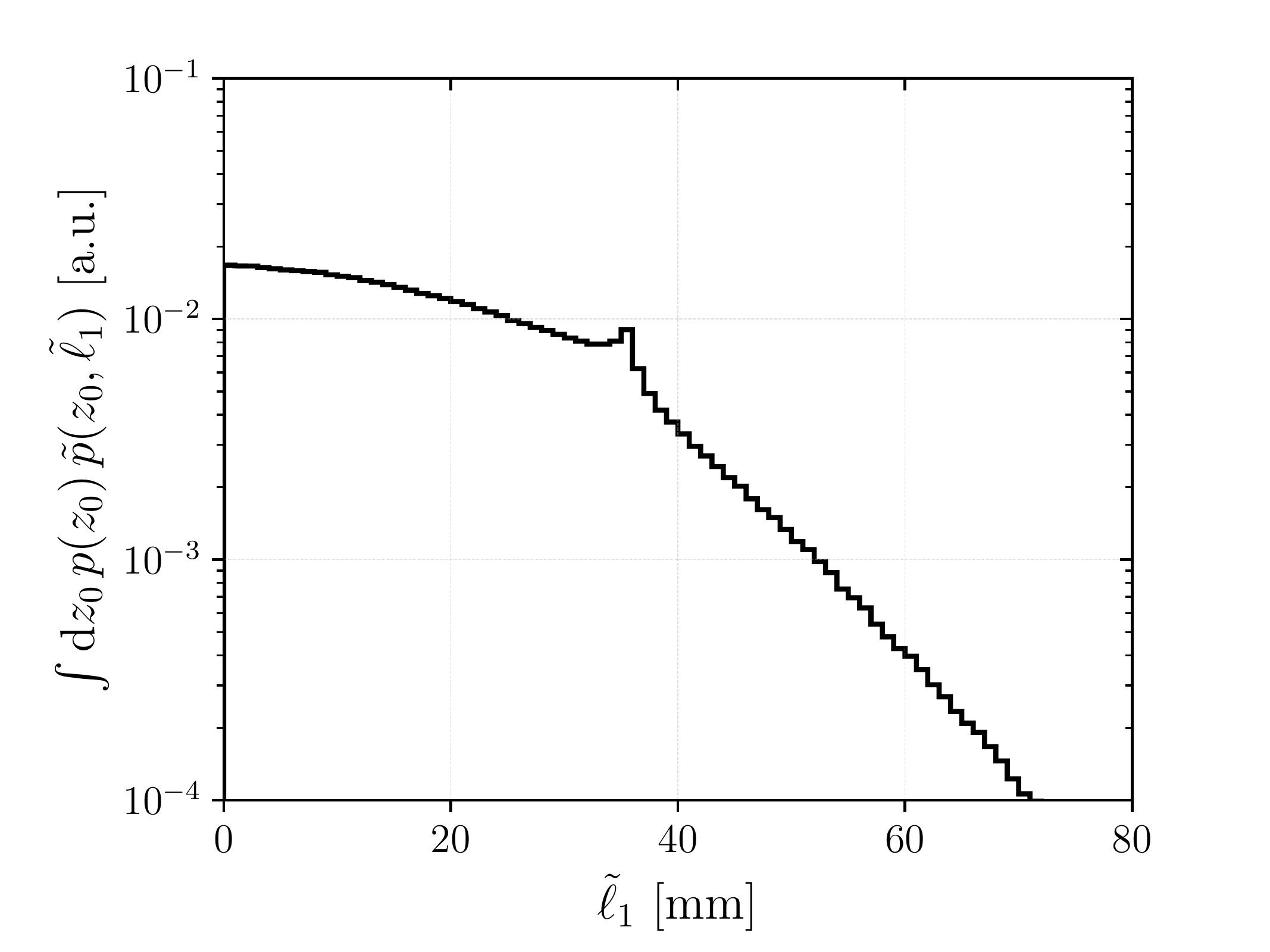}
    \caption{\textbf{Left:} $\tilde p(z_0,\tilde \ell_1)$ for a set of fixed $z_0$. \textbf{Right:} The convoluted distribution over all $z_0$, weighted by $p(z_0)$.}
    \label{fig:distance_distribution_uniform}
\end{figure}

\begin{figure}[!t]
\centering
    \includegraphics[width=0.495\linewidth,trim={1.5cm 0.6cm 2.5cm 1.3cm}, clip]{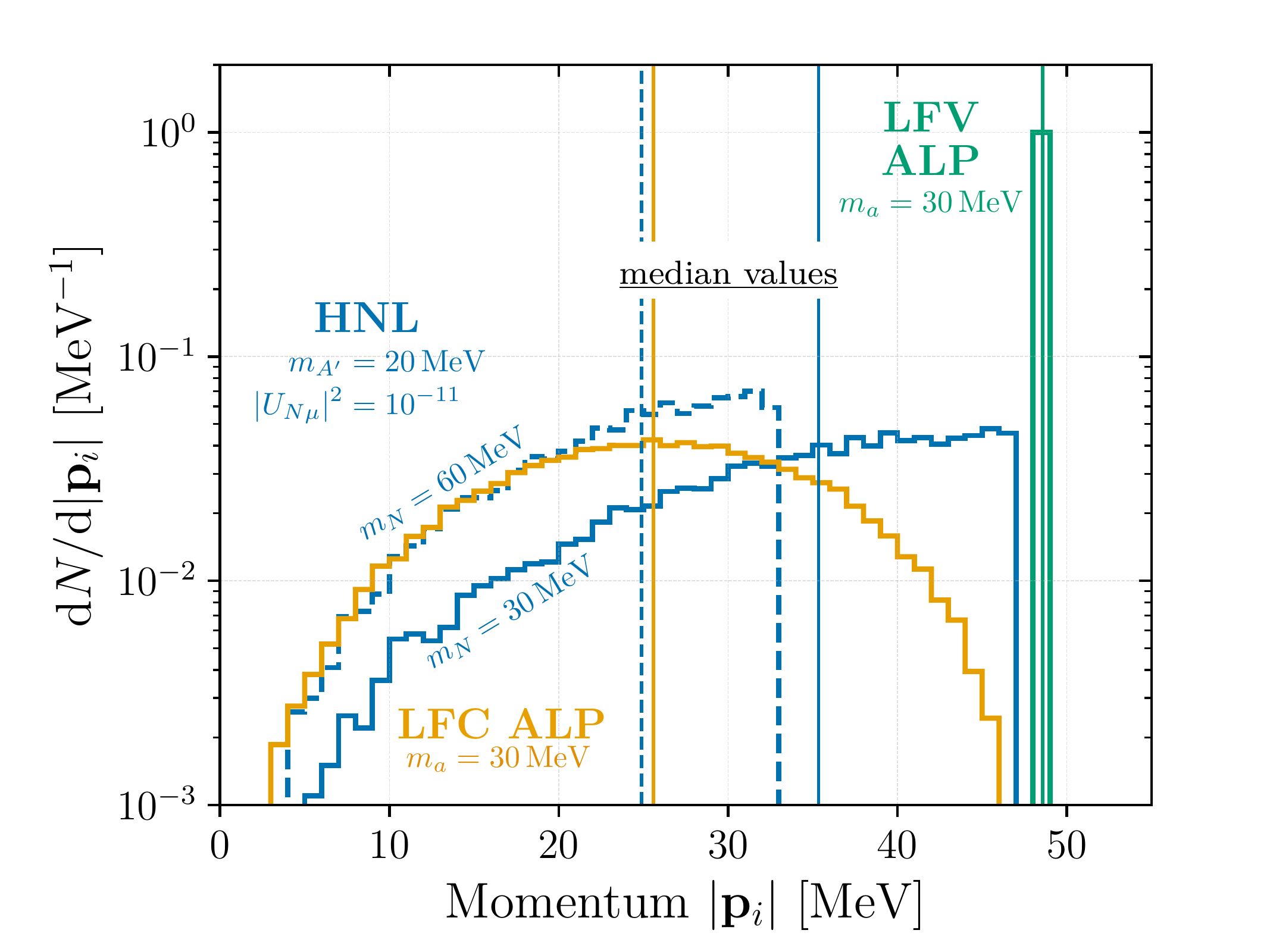} 
    \includegraphics[width=0.495\linewidth,trim={1.4cm 0.6cm 3.2cm 1.1cm}, clip]{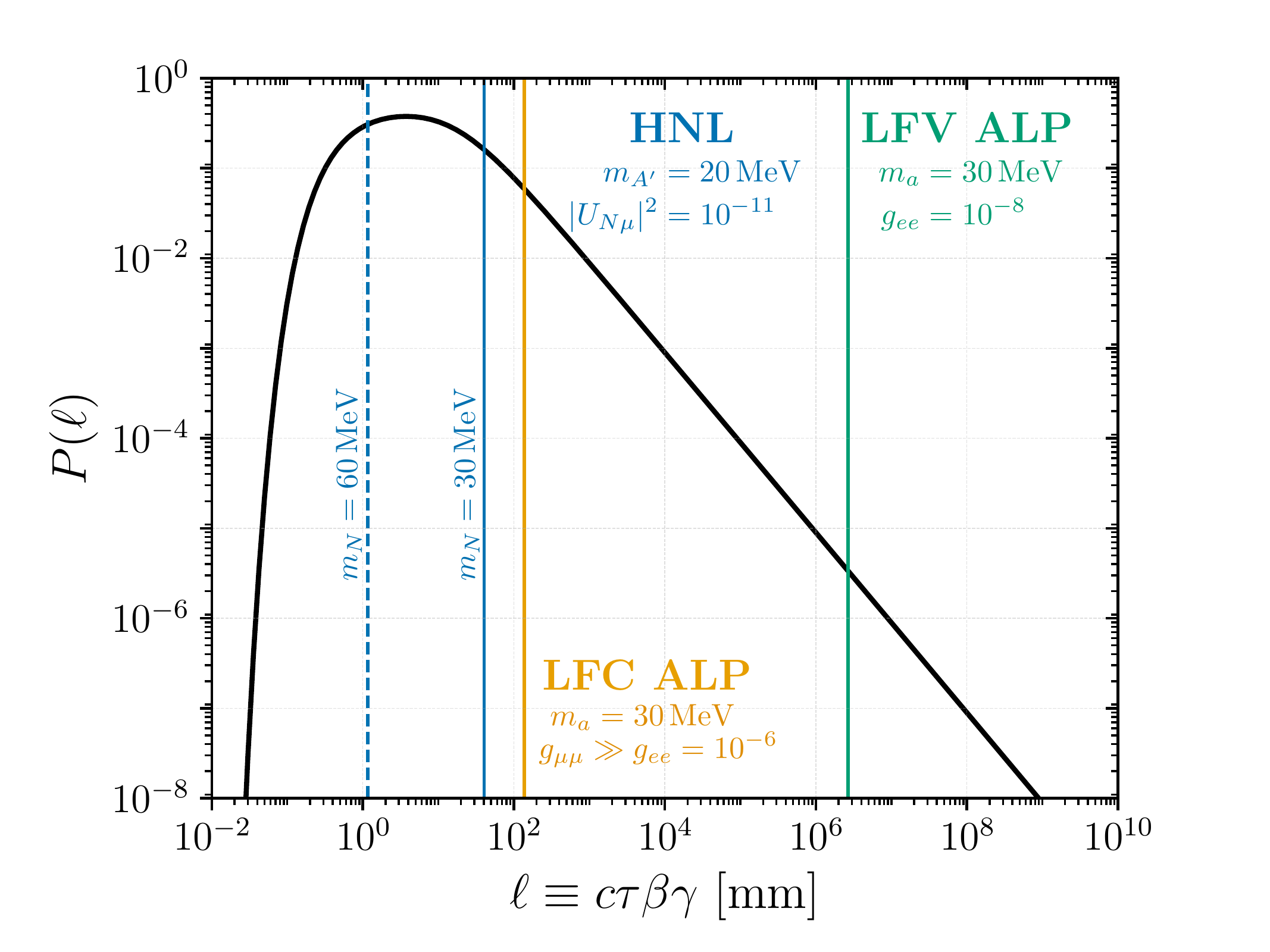}
    \caption{\textbf{Left:} Momentum distributions for representative points from the three models under consideration. Vertical lines indicate the median momentum values for each distribution, which are referenced in the right-hand panel. \textbf{Right:} Probability distribution $P(\ell)$, defined in \cref{eq:weights2}, as a function of the lab-frame proper decay distance. The vertical lines correspond to the $\ell$ values associated with the median momentum values of the model points shown in the left panel.}
    \label{fig:Pell-and-momenta-dists}
\end{figure}

A brute force Monte Carlo sampling is not feasible for over $10^{15}$ muons, and we therefore need to make additional assumptions to compute the $P(\vec p_X,c\tau_X)$ and $\epsilon_i (\{\vec p^i_e\})$ weights efficiently.
Concretely, we find that $X$ can have at most an $\mathcal{O}(1)$ boost (see \cref{fig:Pell-and-momenta-dists}), and as such we can neglect the dependence of $\epsilon_i (\{\vec p^i_e\})$ on the direction of $\vec p_X$. 
Furthermore, if we also neglect the effect of the muon polarization, the muon differential width is spherically symmetric in $\vec p_X$.
This then allows us to replace $P(\vec p_X,c\tau_X)$ in \cref{eq:mastereqappendix} with its average over the direction of $\vec p_X$. 
This function only depends on the the proper decay length
$\ell\equiv\gamma\beta c\tau_X$. With a slight abuse of notation, we will refer to it as $P(\ell)$:
\begin{equation}
    P(\ell) \equiv \frac{1}{4\pi}\int \diff z_0\, \diff\phi  \diff\cos\theta\,  p(z_0)\times A(z_0,\phi,\cos\theta) \times \left[e^{-\ell_0/\ell}  -   e^{-\ell_1(z_0,\phi,\cos\theta)/\ell} \right] \label{eq:weights}
\end{equation}
The lab-frame azimuthal ($\phi$) and polar ($\theta$) angles specify the direction of $\vec p_X$.
The function $A(z_0,\phi,\cos\theta)$ accounts for whether the long-lived particle travels into the interior of the stopping target.
\begin{equation}
    A(z_0,\phi,\cos\theta)=\left\{ \begin{array}{ll} 1 & \text{if the momentum vector points into target}\\
    0 & \text{if the momentum vector points away from the target}\end{array}\right.
\end{equation}
The parameter $\ell_0$ is the spatial resolution of the \muthreee detector, which exceeds the thickness of the target. $\ell_1(z_0,\phi,\cos\theta)$ is the distance along the long-lived particle's trajectory to the point where it exits the target. A priori analytic formulas for $A$ and $\ell_1$ can be obtained, though they are cumbersome. The integral in \cref{eq:weights} can also be evaluated fairly straightforwardly with Monte Carlo, rewriting it as 
\begin{equation}
     P(\ell) =  \int\diff \tilde\ell_1 \left[e^{-\ell_0/\ell}  -   e^{-\tilde\ell_1/\ell} \right] \int\diff z_0\, p(z_0) \times \tilde p(z_0,\tilde \ell_1) \,, \label{eq:weights2}
\end{equation}
with
\begin{equation}
    \tilde p(z_0,\tilde \ell_1) \equiv \frac{1}{4\pi} \int \diff\phi\, \diff\cos\theta A(z_0,\phi,\cos\theta) \times \delta\left[\tilde \ell_1-\ell_1(z_0,\phi,\cos\theta)\right]\,.
\end{equation}
The function $\tilde p(z_0,\tilde \ell_1)$ and its integral weighted by $p(z_0)$ are plotted in Fig.~\ref{fig:distance_distribution_uniform}, while in \cref{fig:Pell-and-momenta-dists} we show the momentum distributions for the long-lived particles for representative parameter points, as well as the resulting function $P(\ell)$ after integration. 

\subsection{Track reconstruction efficiency}

Once an electron/positron is produced in the process, its trajectory will be curled by the magnetic field of the experiment. 
In order for the track to be reconstructed it needs to leave at least four hits in the detector. 
In previous work we estimated that this is the case for $52\%$  of all events for which all three tracks have a transverse momentum of at least $p_T>\SI{12}{\MeV}$~\cite{Knapen:2023iwg}. 
We assume that the same factor holds for displaced tracks, which we consider to be a plausible assumption, as long as we restrict ourselves to displaced vertices in the interior of the target \cite{mu3e:2020gyw}.\footnote{
This assumption is a priori not necessary, as some decays outside the target may also be reconstructed. 
We are however relying on the track reconstruction studies which \muthreee performed for tracks originating from the surface of the target. 
We expect that this is a reasonable approximation of the reconstruction efficiency for tracks which arose from decays inside the target.
The acceptance for tracks from decays outside the target should drop as a function of the distance of the target.
Quantifying this subtle effect is best left to the experts within the \muthreee collaboration, and we therefore choose to not include any such events, which is conservative.}

To compute these efficiencies, we implemented both the scalar and HNL models in {\tt MadGraph}~\cite{Alwall:2014hca} and simulated $10^4$ decay chains for each case of interest. 
We define two different efficiencies, $\epsilon_{3e}$ and $\epsilon_{2e}$, for three and two reconstructed tracks respectively. 
The three-track efficiency is obtained by computing the fraction of simulated events with all three tracks passing the $p_T$ cut (multiplied by the flat 52\% efficiency above). 
The two-electron efficiency is instead obtained by taking the fraction events for which both displaced tracks pass the cut. For simplicity, we assume the same 52\% reconstruction efficiency as for the three track case, which is a conservative estimate.

\begin{figure}[t]
\centering
    {\large \bf Leptophilic ALP} \\
    \hspace{1cm}Lepton flavor violating 2-body \hspace{4cm} Lepton flavor conserving 4-body
    \includegraphics[width=0.495\linewidth,trim={0.25cm 0.2cm 0.2cm 0.25cm}, clip]{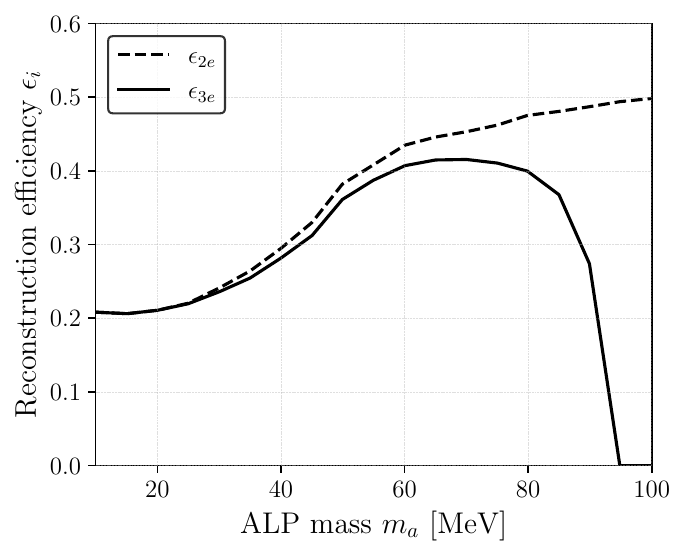}\includegraphics[width=0.495\linewidth,trim={0.25cm 0.2cm 0.2cm 0.25cm}, clip]{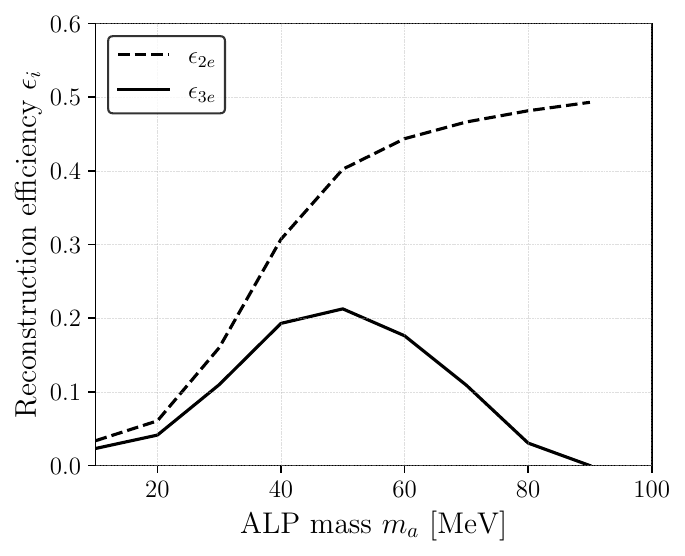} \\[0.3cm]
    {\large \bf Heavy Neutral Lepton}\\
    \hspace{1cm}Long-lived $N$ \hspace{7.5cm} Long-lived $A^\prime$
    \includegraphics[width=0.495\linewidth,trim={1.9cm 0.6cm 2.9cm 1.3cm}, clip]{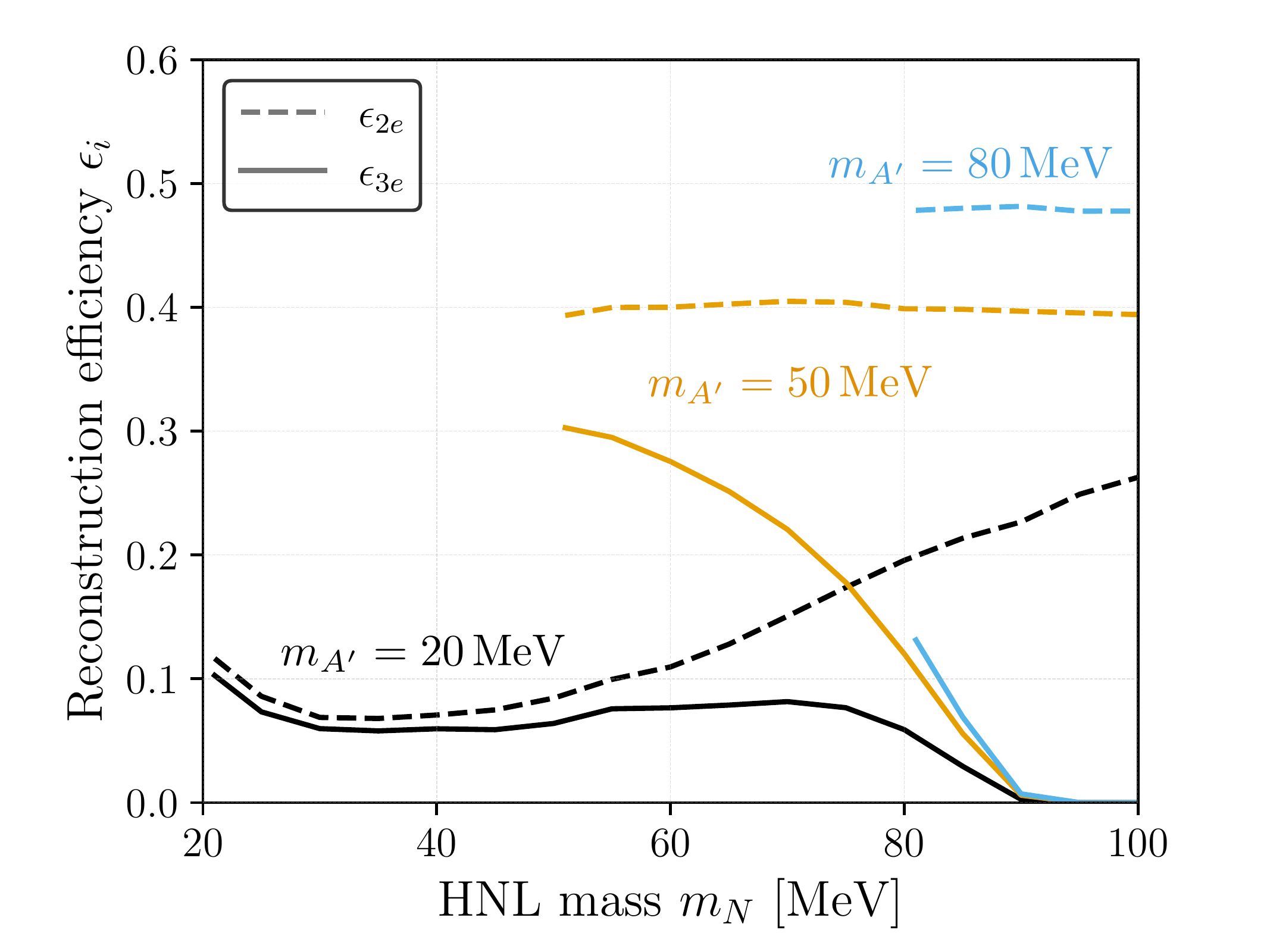}
    \includegraphics[width=0.495\linewidth,trim={1.9cm 0.6cm 2.9cm 1.3cm}, clip]{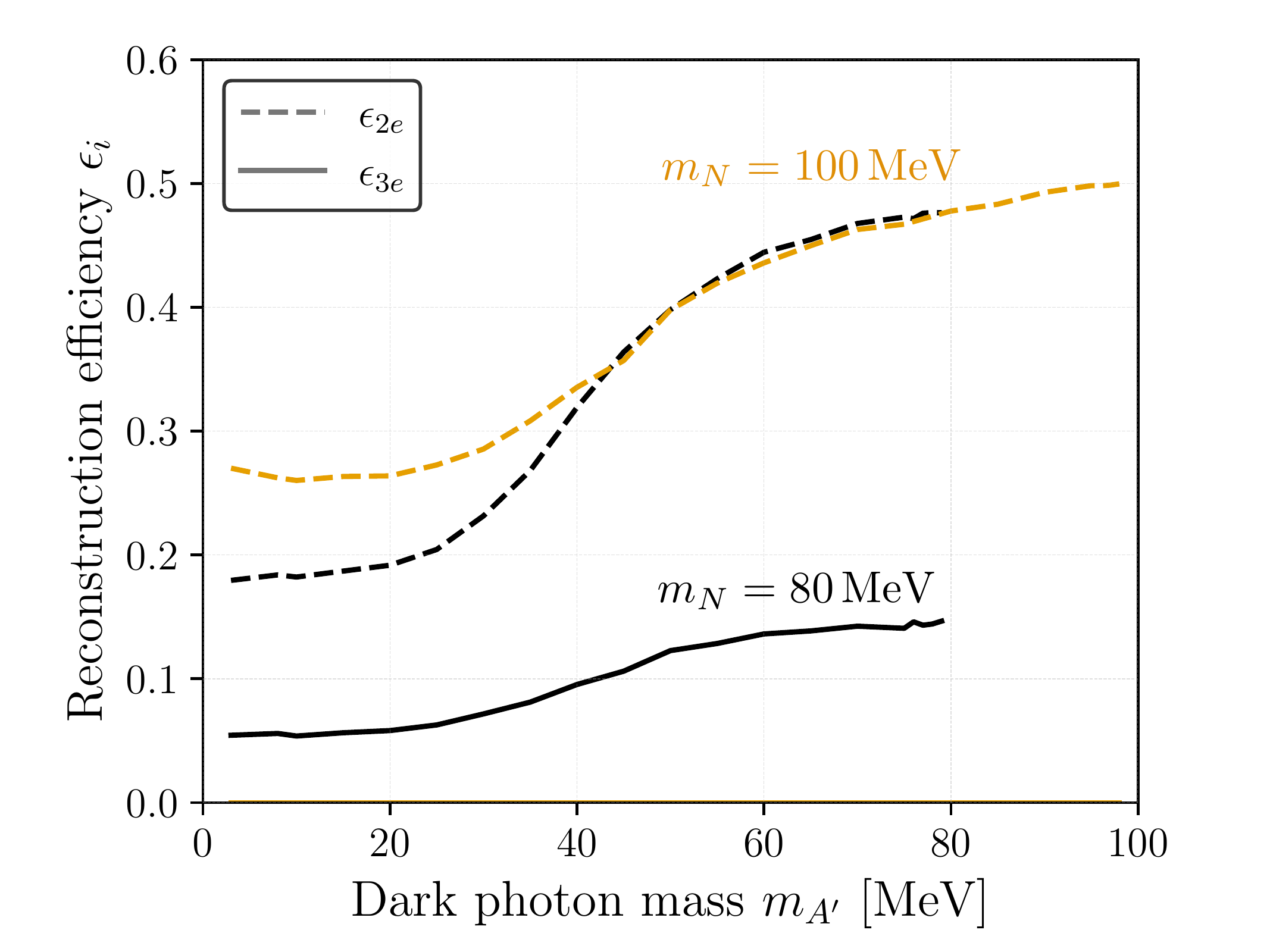}
    \caption{Reconstruction efficiency for $3e$ (solid) and $2e$ (dashed) cases, for the leptophilic ALP model (top), and for the HNL model with prompt $A'$ (bottom left) and prompt $N$ (bottom right), as function of the new particle mass. In the latter model, we show in black, orange and purple the curves obtained by fixing $m_{A'} = 20,50,80$ MeV, respectively. }
    \label{fig:pTacceptance}
\end{figure}

These efficiencies are shown in Fig.~\ref{fig:pTacceptance} for the scalar (left) and HNL (right) benchmark models respectively. In the latter, we show the acceptances for three fixed masses of the light vector boson $A'$, $m_{A'} = 20,50,80$ MeV. 
In all cases, at large masses the $3e$ acceptances goes to zero, as the first electron emitted will have a maximum energy $m_\mu - m_{X}\lesssim \SI{12}{\MeV}$ hence falling below the transverse momentum cut. 
In the same regime however, the two electrons produced by the new particle decay will carry each $E\sim m_{X}/2$, allowing a larger fraction to pass the momentum cut.

%===========================================================================
\section{Secondary Muons at NA62 Proton Beam Dump}
\label{sec:OtherObservables}
%===========================================================================

We consider the secondary muon flux produced at NA62 in dump mode~\cite{NA62:2023qyn}, which has been simulated in Ref.~\cite{Rella:2022len}.
The idea is that the muons could decay in flight to an ALP plus electron, for which we saturate the branching limit $\text{BR}\lp\mu\to e+{\rm inv.}\rp \leq \num{2.6e-6}$ from Jodidio~et.al.~\cite{Jodidio:1986mz}.
This is then followed by a displaced decay of the ALP to an $e^+e^-$ pair, with a proper lifetime $c\tau_a$.

In dump mode, NA62 dumps its proton beam on a Copper (0.8 m) and Iron (2.4 m) target, with a combined length of $L_{\rm tg} = 3.2$ m, followed by the remainder of the beamline with a length of $L_{\rm sp} = 75.8$ m, and by the fiducial decay volume, $L_{\rm dec} = 81$ m. 
The latter ends with the NA62 detector. 
The B1C and B2 magnets downstream from the target are configured to sweep away as many muons as possible, which makes it challenging to properly estimate the contribution of muons decaying in flight after the target.
We therefore only consider muons which decay in flight inside the target; as a result our estimates for the signal yield may be too low by an $\mathcal{O}(1)$ amount.
We estimated the effect of the muon's energy loss in the target~\cite{ParticleDataGroup:2020ssz} before decay, and found this to be a comparably small effect for the energies which would pass the event selection.

Analogous to the NA62 searches in~\cite{NA62:2023qyn, NA62:2023nhs}, we require the electrons and positrons in the final state to pass the cut $E_{e^\pm}\geq10$ GeV~\cite{NA62:2023qyn, NA62:2023nhs}. 
We further assume that both the muons and its decay product will be very forward boosted, and we can assume the angular acceptance is $\sim1$.
Finally, we assume the detector acceptance for the electron-positron pair to be ${\cal A} = 4\%$, analogous to the search in~\cite{NA62:2019eax}. 
The number of signal events can then be estimated by the following integral over the muon energy $E_\mu$
\beq
N_\text{sig.} \approx N_{\rm POT}\times{\cal A} \times \int_{E_\mu^{\rm min}}^\infty {\rm d}E_\mu \frac{{\rm d}f_\mu}{{\rm d}E_\mu} \left[1-\text{exp}\left(-\frac{L_\text{tg}}{E_\mu/m_\mu c\tau_\mu}\right)\right]\left[\text{exp}\left(-\frac{L_\text{sp}}{\ell_a}\right)-\text{exp}\left(-\frac{L_\text{sp}+L_\text{dec}}{\ell_a}\right)\right]\,,
\eeq
where with $c\tau_\mu = 660$ m, $E_\mu^{\rm min}=40$ GeV and $\ell_a\equiv \gamma\beta_a c\tau_a$ the lab-frame decay length of the ALP. 
Here, $\frac{{\rm d}f_\mu}{{\rm d}E_\mu}$ is the probability that a proton produces a secondary muon with energy $E_\mu$, as simulated in~\cite{Rella:2022len}.
Since the scalar is produced by a two body decay, we can estimate $(\gamma\beta)_a \simeq E_\mu/(2 m_a)$. 
Assuming the current NA62 dataset of $1.4\times 10^{17}$ protons on target, we find at most $\sim 1$ expected signal events for any $c\tau_a$ and $m_a$.
When properly accounting for the kinematics of decay chains through a toy Monte Carlo simulation, we find that this number grows to a handful of events for a very narrow range of $c\tau_a$, though probably still marginal for NA62 once backgrounds and more realistic selections are taken into account.
This channel may be interesting however in future facilities with substantially larger data sets, such HIKE or SHiP.

\end{document}